\documentclass[onecolumn,draftcls,12pt]{IEEEtran}
\usepackage{amsmath,amssymb,amsfonts}
\usepackage{textcomp} 
\usepackage{graphicx}
\usepackage{amsmath,amssymb,epsfig} 
\usepackage{varioref}
\usepackage{cleveref}
\usepackage{subfigure}
\usepackage{enumerate}
\usepackage{color}
\usepackage[sort&compress,square,numbers]{natbib}

\usepackage{aliascnt}
\newcommand{\sfE}{\mathsf{E}}

\newtheorem{theorem}{Theorem}
\crefname{theorem}{Theorem}{Theorems} 
\newaliascnt{proposition}{theorem}
\newtheorem{proposition}[proposition]{Proposition}
\aliascntresetthe{proposition}
\crefname{proposition}{Proposition}{Propositions} 
\newaliascnt{lemma}{theorem}
\newtheorem{lemma}[lemma]{Lemma}
\aliascntresetthe{lemma}
\crefname{lemma}{Lemma}{Lemmas} 
\newaliascnt{corollary}{theorem}
\newtheorem{corollary}[corollary]{Corollary}
\aliascntresetthe{corollary}
\crefname{corollary}{Corollary}{Corollarys}

\newtheorem{assumption}{Assumption}
\crefname{assumption}{Assumption}{Assumptions}

\newaliascnt{claim}{theorem}
\newtheorem{claim}[claim]{Claim}
\aliascntresetthe{claim}
\crefname{claim}{Claim}{Claims}

\crefname{figure}{Fig.}{Figures}
\crefname{section}{Sec.}{Sections} 
\newtheorem{definition}{Definition}
\crefname{definition}{Definition}{Definitions}

\crefname{commenv}{Comment}{Comments}
\newtheorem{remark}{Remark}
\crefname{remark}{Remark}{Remarks}
\crefname{footnote}{Footnote}{Footnotes}

\newcommand{\beq}{\begin{equation}}
\newcommand{\eeq}{\end{equation}}
\newcommand{\bea}{\begin{array}}
\newcommand{\ena}{\end{array}}
\newcommand{\bds}{\begin {itemize}}
\newcommand{\eds}{\end {itemize}}
\newcommand{\bdf}{\begin{definition}}
\newcommand{\blm}{\begin{lemma}}
\newcommand{\edf}{\end{definition}}
\newcommand{\elm}{\end{lemma}}
\newcommand{\bthm}{\begin{theorem}}
\newcommand{\ethm}{\end{theorem}}
\newcommand{\bprp}{\begin{prop}}
\newcommand{\eprp}{\end{prop}}
\newcommand{\bcl}{\begin{claim}}
\newcommand{\ecl}{\end{claim}}
\newcommand{\bcr}{\begin{coro}}
\newcommand{\ecr}{\end{coro}}
\newcommand{\bquest}{\begin{question}}
\newcommand{\equest}{\end{question}}

\def\bm#1{\mbox{\boldmath $#1$}}

\newcommand{\avec}{{\bf{a}}}
\newcommand{\bvec}{{\bf{b}}}

\newcommand{\pvec}{{\bf{p}}}

\newcommand{\uvec}{{\bf{u}}}

\newcommand{\xvec}{{\bf{x}}}

\newcommand{\svec}{{\bf{s}}}
\newcommand{\vvec}{{\bf{v}}}

\newcommand{\gvec}{{\bf{g}}}

\newcommand{\hvec}{{\bf{h}}}

\newcommand{\onevec}{{\bf{1}}}

\newcommand{\alphavec}{{\bf{\alpha}}}

\newcommand{\Amat}{{\bf{A}}}

\newcommand{\Imat}{{\bf{I}}}

\newcommand{\Pmat}{{\bf{P}}}

\newcommand{\Nmat}{{\bf{N}}}

\newcommand{\E}{\sfE}

\newcommand{\real}{{\mathbb{R}}}
\newcommand{\comp}{{\mathbb{C}}}
\newcommand{\nat}{{\mathbb{N}}}

\newcommand{\calN}{{\cal N}}

\newcommand{\calC}{{\cal C}}

\newcommand{\0}{{\mathbf {0}}}

\newcommand{\define}{\stackrel{\triangle}{=}}

\def\Qmq{\mathcal{Q}_{\mbox{\begin{footnotesize}-\end{footnotesize}\begin{tiny}$q$\end{tiny}}}}

\def\alphavec{{\mbox{\boldmath $\alpha$}}}

\newcommand{\beqna}{\begin{eqnarray}}
\newcommand{\eeqna}{\end{eqnarray}}

\usepackage[acronym]{glossaries}
\newif\ifisonecolumn
\isonecolumntrue 
\ifisonecolumn
\newcommand{\Dcond}[2]{#2}
\else
\newcommand{\Dcond}[2]{#1}
\fi

\newif\ifisarchive
\ifisarchive

\else

\fi

\usepackage{etoolbox}

\newacronym{jt}{JT}{joint transmission}
\newacronym{rhs}{r.h.s.}{right-hand side}

\newacronym{cran}{C-RAN}{cloud radio access network}
\newacronym{5g}{5G}{fifth generation}
\newacronym{snr}{SNR}{signal-to-noise ratio}
\newacronym{sinr}{SINR}{signal-to-interference-plus-noise ratio}
\newacronym{miso}{MISO}{multiple-input single-output}
\newacronym{mimo}{MIMO}{multiple-input multiple-output}
\newacronym{dmimo}{D-MIMO}{distributed \gls{mimo}}
\newacronym{iid}{i.i.d.}{independent identically distributed}
\newacronym{zf}{ZF}{zero-forcing}
\newacronym{ms}{MS}{mobile station}
\newacronym{mc}{MC}{Monte Carlo}
\newacronym{cdf}{c.d.f.}{cumulative distribution function}
\newacronym{csi}{CSI} {channel state information}
\newacronym{paq}{P\&Q}{precode and quantize}
\newacronym{rvq}{RVQ}{random vector quantization}
\newacronym{bbu}{BBU}{base-band unit}
\newacronym{bs}{BS}{base station}
\newacronym{rrh}{RRH}{remote radio-head}
\newacronym{sud}{SUD}{single-user decoding}
\newacronym{srrh}{S-RRH}{smart \gls{rrh}}
\newacronym{cmi}{CMI}{channel magnitude information}
\newacronym{wlg}{w.l.o.g.}{without loss of generality}
\newacronym{cdi}{CDI}{channel directional information}
\newacronym{mrc}{MRC}{maximum ratio combining}
\newacronym{csit}{CSIT} {\gls{csi} at the transmitter}
\newacronym{dcsi}{D-CSIT} {distributed \gls{csit}}
\newacronym{ccsi}{C-CSIT} {centralized \gls{csit}}
\newacronym{jpm}{JPM} {joint precoding matrix}
\newacronym{jpcu}{JPMCU}{\gls{jpm} computation unit}

\usepackage[normalem]{ulem}

\begin{document}

\title{C-RAN Zero-Forcing with Imperfect CSI:  Analysis and Precode\&Quantize Feedback}
\author{Niv Arad,~\IEEEmembership{Student Member,~IEEE,}
        and Yair Noam,~\IEEEmembership{Member,~IEEE}%

\thanks{N. Arad and Y. Noam are with the Faculty of Engineering, Bar-Ilan University, Ramat-Gan, 5290002 Israel  (e-mail: nivarad44@gmail.com; yair.noam@biu.ac.il). Some of the results reported here appeared in
{\em IEEE ICC 2018}:  ``Precode and quantize channel state information sharing
  for cloud radio access networks,
  pp.~1--6. 
. This research was supported by an Israel Ministry of Science ``Kamin" grant. }
}
\maketitle
\begin{abstract}
Downlink joint transmission by a  cluster of \glspl{rrh} is essential  for enhancing throughput in future cellular networks. This method requires global  \gls{csi} at the processing unit that designs the joint
precoder. To this end, a large amount of \gls{csi}  must be shared between the \glspl{rrh} and that unit. This paper proposes two contributions. The first is a new upper bound on the rate loss, which implies a lower bound on the achievable rate for an \glspl{rrh}-cluster employing joint \gls{zf} with incomplete \gls{csi} with \gls{sud} at the \glspl{ms}. The second contribution, which follows insights from the bound, is a new \gls{csi} sharing scheme that drastically reduces the significant overhead associated with acquiring global \gls{csi}  for joint transmission. In a nutshell, each \gls{rrh} applies a local precoding matrix that  creates low-dimensional effective channels that can be quantized more accurately with fewer bits, thereby reducing the overhead of sharing \gls{csi}. In addition to the \gls{csi}  sharing-overhead, this scheme reduces the data rate delivered to each \gls{rrh} in the cluster.  
 \end{abstract} 
\section{Introduction}
\Gls{jt} is a key   for enhancing spectrum utilization in wireless communication networks. We consider downlink  \gls{jt}, in which adjacent \glspl{rrh} form a cluster serving multiple
\glspl{ms}; a setup also known as \gls{dmimo}. The idea is to transform  interference between adjacent cells  into useful signals. To do so,  \glspl{rrh} in the cluster must share both the data and \gls{csi}. Such a high level of cooperation is the main obstacle to exploiting
the vast potential of \gls{jt} in practice. 
\Gls{cran} architecture facilitates the  high level of \gls{dmimo}  cooperation  via a centralized \gls{bbu} pool connected via high data-rate links, dubbed \(fronthaul\), to a large number of \glspl{rrh} \cite{checko2015cloud}. That \gls{bbu} performs  all digital processing centrally, which is excellent for \gls{jt}. 
 However,  \gls{jt}  requires  {ultra high-rate} data sharing and high-rate, low-latency sharing of \gls{csi}  between the \gls{bbu} and all \glspl{rrh}, which the fronthaul does not always support. This obstacle has motivated research on reducing 
fronthaul data-rate (see. e.g., \cite{dai2013sparse,zakhour2011optimized}) and for the 
introduction of a more functional \gls{rrh}, dubbed \gls{srrh}, that 
carries out some of the digital  processing \label{PageMarkSRRH} \cite{Larsen2019}.\footnote{\label{FNoteSRRH} We use the term \gls{srrh} for radio units in
 \gls{cran} architectures with functional-splitting at higher-layers (see, e.g., \cite{Larsen2019}); i.e.,  higher than the layer in the original \gls{cran} concept (dubbed today Splitting 8), where the radio unit performs only radio functions of converting baseband IQ symbols into analog signals and vise verse  \cite{Larsen2019}. On higher-layer splittings, the radio unit is more functional; e.g., in Splitting 7-1, 7-2, and 7-3, the radio unit carries out low-PHY functionalities, wherein Splitting 6 it performs all PHY functionalities.
 The \gls{bbu} functionality is divided between a central unit (CU), a distribution unit (DU), and the radio unit that execute higher to lower layer functions. The DU is typically close to the radio
unit.} 

 {\gls{dmimo} setups differ from one another in the type of  \gls{csit}. 
In the first type, dubbed {\em \gls{ccsi}}
\cite{DeKerret2013b}, each \gls{srrh} 
sends its \gls{csit} to the \gls{bbu}. The latter thus has a single estimate of the global \gls{csit}, from which it
calculates the \gls{jpm}. Finally, the \gls{bbu}  feeds each  \glspl{srrh} its corresponding
 \gls{jpm} sub-block perfectly.}  
 {\label{PageCsiTypes}In  another type of  \gls{csit},  dubbed {\em  \gls{dcsi}} \cite{DeKerret2013,Li2020c,DeKerret2014}, no single entity calculates the \gls{jpm} based on a single global-\gls{csit} estimate. Instead, each \gls{srrh} broadcasts its local \gls{csit} to  other \glspl{srrh} (e.g., via a low-latency wireless broadcast channel), then estimates the \gls{csit} locally, leading to a different global-\gls{csit} estimate for each \gls{srrh}.
Finally, the \gls{srrh} calculates its \gls{jpm} from its locally known global-\gls{csit}.}

\begin{figure}[h!]
        \centering
        \includegraphics[width=0.7\textwidth]{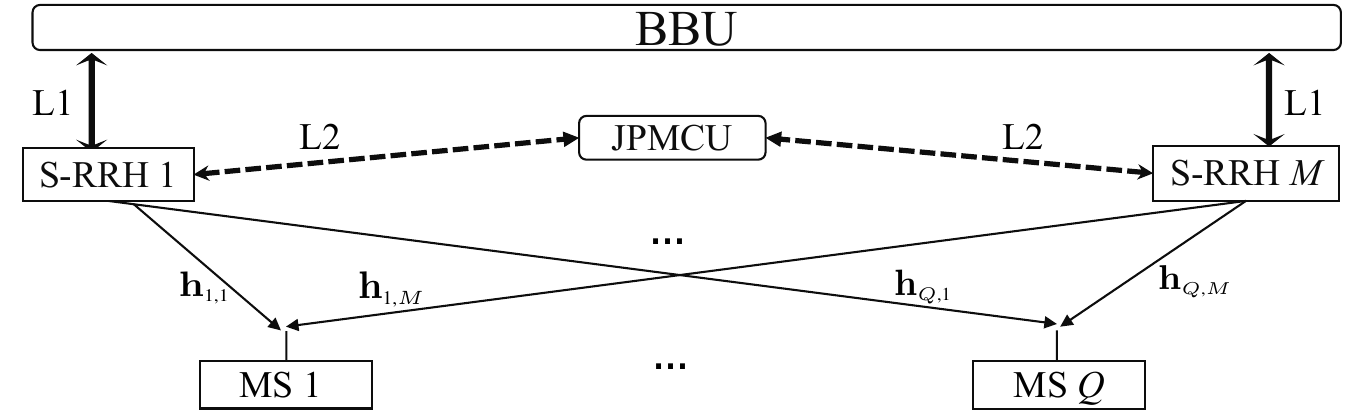}
        \caption{System model. {Link L1 interconnects the \glspl{srrh} to higher-level \gls{cran} functions (see \cref{FNoteCRANExp} for details). The joint precoding-matrix
computation unit (\gls{jpcu}),  located close to the \glspl{srrh}, is connected
via the low-latency, rate-limited link L2.}}
        \label{FigSystemSetupICSI}
\end{figure}

{We consider a different setup in which  a centralized computation unit, having global \gls{csi}, calculates the  \gls{jpm}. Explicitly, each \gls{srrh} sends
its \gls{csit} to that unit, henceforth dubbed \gls{jpcu}, via a low-latency albeit rate-limited link  as depicted in  \cref{FigSystemSetupICSI}.\footnote{\label{FNoteCRANExp} {Here are examples of practical \gls{cran} configurations where our setup (cf. \cref{FigSystemSetupICSI}) is suitable. The first is where enhanced radio units (as in functional splittings 7-1,7-2,7-3 and 6 cf.  \cite{Larsen2019}) are connected via wireless fronthaul to a distribution unit (DU), which, with the central unit (CU), constitutes the \gls{bbu}. In this case, the \gls{jpcu} is at the \gls{bbu} (cf. \cref{FigSystemSetupICSI}), and L1 and L2 are the same link; i.e., the fronthaul, which is wireless, hence rate limited.  Another scenario is in functional splittings where the DU and CU are physically separate and connected via a mid-haul link (cf.  \cite{Larsen2019}), which typically has high latency that does not support \gls{jpm} calculation. In this scenario, the \gls{bbu} in \cref{FigSystemSetupICSI} represents the CU, each \gls{srrh} is a DU, and L1 is the mid-haul link. Then, to facilitate \gls{jt}, it is sufficient to have a rate-limited L2 link between \gls{jpcu} and  \gls{srrh}s, acting as Xn \cite{Larsen2019}.}}
  We note  that we consider only  \gls{csi}-quantization errors while neglecting \gls{csi} errors due to latency (outdated CSI). Upon receiving global \gls{csit}, the \gls{jpcu} calculates the \gls{jpm}.
However,  unlike the C-\gls{csit} setup (where the error is only in the \gls{csit} at the BBU),
the \gls{jpcu} does not send each  \gls{srrh}  its corresponding submatrices
perfectly but instead sends a quantization. 
 The proposed  setup is similar to the \gls{dcsi} in that the employed precoding matrix   contains errors compared to that of the centralized design. The difference is in the error type. While in \gls{dcsi}, the additional \gls{jpm} error (compared to \gls{ccsi}) follows from  independent \gls{csit}-errors at each \gls{srrh}; in the proposed scheme, that error is due to the quantization of the centralized \gls{jpm}.}

{\Gls{jt} under \gls{ccsi} and \gls{dcsi} is a well-studied topic. The  single transmitter case with \gls{ccsi} was first studied in \cite{jindal2006mimo}, preceded by others (see, e.g. \cite{Caire2010Multiuser}), and extended to the  multi-transmitter case with \gls{ccsi} and \gls{dcsi} in \cite{makki2012throughput, yu2012novel,dai2013sparse, zakhour2011optimized,DeKerret2013,DeKerret2014,jaramillo2015coordinated, Zhang2013a,Sanguinetti2016,Li2018,Li2020c,Pan2017,Pan2017a,Pan2019}. We focus on the case  where the main factor is the limited link between \glspl{srrh} in the cluster and the \gls{jpcu}.\footnote{Explicitly, we assume that each \gls{srrh} estimates the channels between it and all \glspl{ms}, and we neglect that estimation error. This assumption may be reasonable if  there is channel reciprocity, as in time division duplex, where the channel estimation error at the \glspl{srrh} is negligible compared to the quantization error in sending these channels to the \gls{jpcu}.} Here, in addition to reducing the fronthaul data rate,  we deal with another  issue; namely \gls{jt} with imperfect \gls{csi}.}

 The paper presents two contributions. The first is a new upper bound on the rate-loss, where the \gls{jpcu} sets the overall joint-\gls{zf} precoding matrix using imperfect \gls{csi} (cf. \cref{FigSystemSetupICSI}), compared to perfect \gls{csi}, {where \gls{csi} errors are due to quantization}. That upper bound yields a lower bound on the achievable rate.
We assume  that each \gls{srrh} quantizes its local \gls{csi} using \gls{rvq}  \cite{santipach2005signature}. {\label{PageSimilarBounds}Similar bounds for the  broadcast channel and \gls{dmimo} with imperfect \gls{csi}  appear in \cite{jindal2006mimo,Caire2010Multiuser} and \cite{Zhang2013a,Sanguinetti2016,Li2018,Li2019,Wang2020}, respectively, all of which consider \gls{ccsi}. As discussed above, the proposed bound here differs from the latter bounds due to the \gls{jpm} quantization, which does not exist in the \gls{ccsi}. Furthermore, in  \cite{jindal2006mimo,Caire2010Multiuser}, the overall channel to each terminal is quantized
as a haul, whereas here, in sub-blocks.  This sub-block
quantization induces an entirely different \gls{csi} error distribution
leading to a distinct bound. Moreover, \cite{Zhang2013a,Sanguinetti2016,Li2018,Li2019,Wang2020} consider the large system regime, whereas the analysis here does not.\footnote{The large system regime is where the number of \gls{rrh}-antenna and the number of terminals grows to infinity, while their ratio approaches a nontrivial limit. In this regime, it is possible to use random matrix theory.}   Finally,  \cite{Li2018,Wang2020} deal with channel impairment due to pilot contamination, whereas in this paper, the error is due to \gls{csi} quantization. Another relevant rate-loss bound is \cite{DeKerret2013}, which, unlike here, considers the \gls{dcsi} setup, which is different as discussed above. Moreover, the bound  \cite{DeKerret2013} differs from the proposed
bound because it assumes single-antenna transmitters and considers the high-\gls{snr} regime. A recent bound under no such assumption for the \gls{dcsi} setup appears in \cite{Li2020c}. However, beyond the \gls{dcsi}, the latter bound considers the large system regime, whereas  the proposed bound does not.}

The second contribution is a new scheme for precoding and  \gls{csi} sharing, dubbed \gls{paq}, which has two advantages. First, it reduces the number of \gls{csi} quantization bits transferred on the L2-link (cf. \cref{FigSystemSetupICSI}) between the \glspl{srrh} and the \gls{jpcu}. The second advantage is that it reduces the overall fronthaul data rate between the  \glspl{srrh} and the \gls{bbu}. {\label{CsiSchemesComparison} There are different approaches for reducing \gls{jt} \gls{csi}-overhead. One method, used for uplink \gls{jt}, compresses the \gls{csi}  delivered to the \gls{jpcu}  \cite{kang2014joint}. Other techniques are robust (to inaccurate \gls{csi}) precoding \cite{wang2017robust,lakshmana2016precoder} and compressive \gls{csi} acquisition  \cite{shi2014csi}. De Kerret and Gesbert  \cite{DeKerret2014} proposed spatial \gls{csit} allocation policies maximizing the generalized degrees of freedom. The latter study indicates that TX cooperation should be limited to a specific finite neighborhood around each TX. Sanguinetti et al.  \cite{Sanguinetti2016} designed linear precoders that minimize power consumption under a target-rate constraint and further analyzed its performance in the large system regime. Pan et al.  \cite{Pan2017} proposed a user selection algorithm and joint precoding design, which reduces implementation complexity under perfect \gls{csit} or pilot-contamination  \cite{Pan2019}. A key distinguishing characteristic of this scheme is that it applies front-end precoding matrices at the \glspl{srrh} before \gls{csi} quantization. These matrices aim at improving \gls{csi} accuracy at the \gls{jpcu}. Each \gls{srrh} autonomously calculates and applies a matrix, based on its local \gls{csi}, whereby creating an effective channel of lower dimensionality that can be quantized more accurately. These channels are then quantized and sent to the \gls{jpcu}, which in turn calculates a joint precoding matrix and feeds it back to the \glspl{srrh}. We show,  theoretically and numerically that this scheme significantly increases the network throughput compared to the standard scheme, in which each \gls{srrh} quantizes its local \gls{csi} and feeds it back to the \gls{jpcu}. This performance gain remains for a wide range of \gls{csi} quantization bits and SNR. Equally important, the proposed feedback scheme significantly reduces  the data load on the fronthaul connecting the \glspl{srrh} and the \gls{bbu}.}

The organization of the paper is as follows. Sec. \ref{SecSystem Model} introduces the system model. In Sec. \ref{SecAnalysisResidualInterference}, a new upper bound on the rate-loss is presented. Sec. \ref{HierarchicalCoMPDescription} describes the \gls{paq} \gls{csi} sharing scheme and its benefits while Sec. \ref{SecPandQPanalysis} analyzes its performance theoretically. Simulation results are given in Sec. \ref{SecNumericalResults} and conclusions in Sec. \ref{SecConclusions}.

{\em Notation}: Boldface lower (upper) case letters denote vectors
(matrices). $(\cdot)^{*}$ and $(\cdot)^{\dagger}$ denote the conjugate
and the conjugate transpose operations, respectively. 
Moreover, $\odot$ and $\otimes$ mark the Hadamard
 and Kronecker products, respectively.   Let    $\avec$, be a vector, then $\bar\avec$ denotes its normalized version; i.e.,   $\bar\avec=\avec/\Vert \avec\Vert.$ Also,    $\angle \langle \avec,\bvec\rangle$ marks the angle between   $\avec$ and the vector $\bvec$.  In addition, let ${\cal Q}$ be a set and $q\in {\cal Q}$, then $\mathcal{Q}_{\mbox{\tiny-}q}={\cal Q} \setminus \{q\}$.  Consider $ \{\Amat_q\}_{q=1}^Q,$ where
        $\Amat_{q}\in \mathbb{C}^{N \times M}$,  then  $\Amat= {\rm blockdiag}(\Amat_1,\ldots,\Amat_Q)\in \mathbb{C}^{Q N\times
                Q M } $  denotes the  block diagonal matrix, whose $q$th diagonal-block is equal
        to  $ \Amat_q$; i.e.,  $ [\Amat]_{(q-1) N+n,(q-1) M+m}=[\Amat_{q}]_{n,m} $  $\forall $ $1\leq q\leq Q$,   $1\leq n\leq N$, $1\leq m\leq M$  and is equal to
        zero otherwise, where $[A]_{l,m}$ denotes the $(l,m)$ entry of $\Amat$. Let \(\mathcal{H} = \rm{span} ({\hvec}_1,\ldots,{\hvec}_M)\), 
then  \(\Pmat_{\mathcal{H}}\), \(\Pmat^{\perp}_{\mathcal{H}}\)
 denote the projection matrices into  space spanned by  \(\mathcal{H}\) and
into  its 
orthogonal complement, respectively. Also, \(\chi_{{\cal A}}(x)\) represents the indicator function; that is,  \(\chi_{{\cal A}}(x)=1\) if \(x \in {\cal A}\) and \(0\) otherwise, $\Imat_{N}$ denotes an $N \times N$ identity matrix and $\onevec_{N}, {\mathbf 0}_{N}$ denote an $N\times 1$ vector of ones, and zeros, respectively. Finally, we use $\log$ for the base 2 logarithm.
 
\section{System Model}
\label{SecSystem Model}
Consider a cluster of   \(M\) \glspl{srrh},  each with \(N_t\) antennas, that jointly serve \(Q\) single-antenna \glspl{ms}, as depicted in Fig. \ref{FigSystemSetupICSI}. We denote the set of \glspl{srrh} \(\{1,...,M\}\) by \({\cal M}\) and the set of \glspl{ms} \(\{1,...,Q\}\) by \(\cal {Q}\). Assuming flat fading channels, the \emph{downlink} signal, observed by\ \gls{ms}-\(q\), is given by
\begin{IEEEeqnarray}{rCl}
        \label{SignalModel1}
y_{q}=\sum_{m=1}^{M} \hvec^{\dagger}_{q,m}\xvec_{m}+n
_{q}
\;, \quad\forall q\in\mathcal{Q}
\end{IEEEeqnarray}
where \(n_q\) is an additive, proper-complex Gaussian noise \(n_q \sim \calC \calN(0,\sigma^2_{n})\), \(\xvec_{m}\in\comp^{ N_t \times 1}\) is the   signal transmitted by \gls{srrh}-\(m\); \(\hvec_{q,m}\in \comp^{ N_t \times 1}\)  is the channel between \gls{srrh}-\(m\) and \gls{ms}-\(q\). We further denote
\begin{IEEEeqnarray}{rCl}
 \label{eq_define hq}\hvec_{q}\define[\hvec^{\dagger}_{q,1},...,\hvec^{\dagger}_{q,M}]^{\dagger} \in\comp^{ M N_t \times 1}.
\end{IEEEeqnarray}
  \begin{assumption}
          The channels are  Rayleigh, \gls{iid} block-fading.\footnote{\label{FNoteBlockFading}We use the standard definition of a block-fading channel (see \cite{tse2005Fundamentals}, Ch. 5.4); that is, a channel that remains constant during a particular time block, dubbed coherence time, which is much shorter than the code block-length. This channel is drawn randomly at each coherence time and forms an ergodic sequence over time.
The block-fading assumption  also applies under interleaving (\cite{tse2005Fundamentals} Ch. 5.4).} Moreover, we assume large-scale fading (e.g., pathloss and shadowing effects), expressed by an attenuation factor $\alpha_{q,m}$. Explicitly, the channel
        $\hvec_{q,m}\sim {\cal CN}(\bm{0}_{N_{ t}},{\alpha_{q,m}}
\Imat_{N_{ t}}), \forall
q\in{\cal Q}, m \in \cal M $  varies at each coherence time, whereas  $\alpha_{q,m}$ remains constant 
        during the entire
codeword.           
 \label{ass_channelModel}
        \end{assumption}
 
\begin{assumption}
 \label{ass_power_constraint}
  We assume a practically
oriented  short-time power constraint $P_{\rm max}$ for each \gls{srrh}; i.e., $\E\left\{\Vert\xvec_m\Vert^2\vert U \right\}\le P_{\max},\;\forall m\in \mathcal{M}$ for every coherence-time, where \(U\) is the overall instantaneous-\gls{csi}. We further assume a linear precoding scheme in which
\(  \xvec_m=\sum_{q\in\mathcal{Q }}s_{q}\pvec_{q,m} \), where \(s_{q}\in\comp\) is the information-bearing
signal intended to \gls{ms}-\(q\) and  \(\pvec_{q,m} \in \comp^{ N_t \times 1} \) is the precoding vector from \gls{srrh}-$m$ to MS-$q$. Finally \(s_1 ,...,s_{q}\) are assumed \gls{iid} and \(s_q \sim \calC \calN(0{,P_{q}})\).
\end{assumption}

We focus on a fully cooperative multi-cell system; thus, the joint downlink transmission can be conveniently 
modeled as a large \gls{miso} broadcast channel  with \(M N_t\) transmitting antennas such that  the signal observed by \gls{ms}-\(q\) is
\begin{IEEEeqnarray}{rCl}
\label{Signal-at-User-q}
y_{q}=\hvec^{\dagger}_{q}
\pvec_{q}s_{q} + \sum_{j\in
\mathcal{Q}_{\mbox{\tiny-}q}}\hvec
^{\dagger}_{q} \pvec_{j}
s_{j} + n
_{q}
\;, \quad\forall q\in\mathcal{
Q}
\end{IEEEeqnarray}
 where  $\E\{\vert s_{q}\vert ^{2}\} = P_{q}$, $ \Vert\pvec_{q}\Vert^{2}=1$ and $\pvec_{q}$  is the overall joint beamforming vector designated for \gls{ms}-\(q\); i.e.,
\beq
{\pvec_{q}}\define[\pvec_{q,1 }^{\dagger},...,\pvec_{q,M}^{ \dagger}]^{\dagger}
\in\comp^{M N_t \times 1}.
\eeq
We assume channel reciprocity (such as in time division duplex) and consider   \gls{sud}; i.e., each \gls{ms} treats the interfering
        signals as noise. Therefore, every \gls{srrh} estimates the channels  between it and each \gls{ms} served by the cluster.  

\begin{assumption}
        \label{ass_long_term_channel}
The long term channel characteristics
are locally known at each \gls{srrh} and 
globally known at the \gls{jpcu}; i.e., for each $m\in{\cal M},$ \gls{srrh}-$m$ knows $\{\alpha_{q,m}\}_{
        q\in{\cal Q}}$ whereas the \gls{jpcu} knows
$\{\alpha_{q,m}\}_{
        m\in {\cal M},q\in{\cal
                Q}}$. Since these parameters are conveyed to the \gls{jpcu} only once, we neglect the associated overhead on the
L2-link  (cf. Fig. \ref{FigSystemSetupICSI}). 
Moreover, for simplicity, we assume
that each 
\gls{srrh}-\(m\) has perfect
local  \gls{csi} \(\{\hvec_{q,m}\}_{q\in
        \mathcal{Q}}\); i.e., no estimation
errors. 
\end{assumption}
\begin{assumption}
     Since the L2-link is rate limited, \gls{srrh}-$m$ quantizes its \gls{csi} and sends the indices of the quantization codewords $\{c_{q,m}\}_{q\in {\mathcal Q}}$ with an overall number of $B_{}$ bits to the \gls{jpcu}. Upon receiving all the  codewords \(U = \{c_{q,m}\}_{q\in {\mathcal Q},m\in{\mathcal M}}\), the \gls{jpcu} estimates \(\hvec_q, \ \forall q\in \mathcal{Q}\) as 
\beq
\label{eq_define hq hat}
\hat\hvec_q \define [\hat\hvec_{q,1 }^\dagger,...,\hat\hvec_{q,M}^ \dagger]^\dagger \in\comp^{ M N_t \times 1}
\eeq
\label{ass_quatnization}
where \(\hat\hvec_{q,m}\) is the estimate of \(\hvec_{q,m}\), \(\forall q\in \cal{Q}\), \( m\in \cal{M}\). 
\end{assumption}

For now, we do not restrict ourselves to a particular quantization or estimation method. Henceforth, we refer to this procedure as the
 \textit{standard} \gls{csi} feedback scheme. Based on \(\{\hat \hvec_{q}\}_{q\in\mathcal{Q}} \), the \gls{jpcu} calculates  the overall joint precoding matrix as follows
\begin{IEEEeqnarray}{rCl}
\label{eq_ZFP}
\pvec_q \define \Nmat_q \frac{( \hat\hvec^{\dagger}_q \Nmat_q)^{\dagger}}{\|\hat\hvec^{\dagger}_q\Nmat_q\|}
\;, \quad\forall q\in\mathcal{Q}
\end{IEEEeqnarray}
where the columns of  \(\Nmat_q \in \comp^{MN_{t} \times MN_t - (Q - 1)}\) form an orthonormal basis for the null space of  $\{\hat{ \hvec}_{j}\}_{j\in\mathcal{Q}_{\mbox{\tiny-}q}}$. Henceforth, we refer to this scheme as \gls{zf} beamforming. After setting $\pvec_q,\forall q\in \mathcal{Q}$, the \gls{jpcu} quantizes it and feeds each \gls{srrh} with its corresponding components. 
\begin{assumption}
        \label{ass_quatnization_back}
 For each $m$, the \gls{jpcu} quantizes  \(\{\pvec_{q,m}\}_{q\in\mathcal{Q}}\) with  overall $B_{}$ bits and then sends to \gls{srrh}-$m$. The corresponding estimate at \gls{srrh}-\(m\), is denoted by \(\{\hat\pvec_{q,m}\}_{q\in\mathcal{Q}}\).
\end{assumption}
 Because \(\pvec_q\) is  orthogonal to \(\{\hat \hvec_j\}_{j\in\mathcal{Q}_{\mbox{\tiny-}q}} \) rather than \(\{\hvec_j\}_{j\in\mathcal{Q}_{\mbox{\tiny-}q}}\), there is a performance loss compared to the case of perfect \gls{csi}  due to residual
interference, even if $\hat \pvec_{q,m}$ is quantized without errors.
For simplicity and analytical tractability, we assume that the data signals $s_q, q\in{\cal Q}$ are delivered to the \glspl{srrh} without errors.


\section{Downlink CRAN-JT:  Performance Analysis for \gls{zf} with imperfect \gls{csi} }
\label{SecAnalysisResidualInterference}

This section introduces a new upper bound on the throughput degradation under limited \gls{csi} compared to perfect \gls{csi}.
For simplicity and analytical  tractability, we assume that  the  \gls{cmi}  \(\Vert\hvec_{q,m} \Vert\), \(\forall q\in \cal{Q}\), \( m\in \cal{M}\) is perfectly conveyed to the \gls{jpcu}. Moreover,  the \gls{cdi}   \(\bar\hvec_{q,m}\define\hvec_{q,m}/\Vert \hvec_{q,m} \Vert\) is quantized separately using \gls{rvq} \citep{santipach2005signature},\footnote{In \gls{rvq}, the codebook is generated randomly from a uniform distribution over the unit sphere.}  with independent codebooks for every \(q,m\).
We assume the same  about \(\{\pvec_{q,m}\}_{q\in\mathcal{Q}, m\in\mathcal{M}}\). We now review some of the properties of \gls{rvq}.  Let  $\hat{\bar \hvec}_{q,m}$ be the output of \gls{rvq} with $b$ bits. Then,  \begin{IEEEeqnarray}{rCl}\label{eq_DecompositionJindalhBar}
        \bar \hvec_{q,m}=\sqrt {1-Z_{q,m}}\hat{\bar\hvec}_{q,m}+\sqrt{Z_{q,m}}\svec_{q,m}
\end{IEEEeqnarray} where $\svec_{q,m}$ is a random vector uniformly distributed on the unit sphere of the null space of
$\hat{\bar{\hvec}}_{q,m}$, and $Z_{q,m}$  is a random variable, independent of $\svec_{q,m}$, and distributed as the minimum of $2^b$ beta $(N_{ t}-1,1)$ random variables \cite{jindal2006mimo}.
Under the assumption of perfect  \gls{cmi},
the \gls{jpcu} use
\begin{IEEEeqnarray}{rCl}
        \hat{\hvec}_{q,m}=\Vert\hvec_{q,m}\Vert\hat{\bar{\hvec}}_{q,m} \ , \label{eq_ Expression for the estimated vector}
\end{IEEEeqnarray}
as the estimate of  $ \hvec_{q,m}$.\footnote{Note that $\E\{\hvec_{q,m}\big\vert\Vert{\hvec_{q,m}}\Vert,\hat{\bar{\hvec}}_{q,m}\}$, the minimum mean square error estimate of $\hvec_{q,m}$, is  given by $\E\{\sqrt{1-Z_{q,m}}\}\Vert\hvec_{q,m}\Vert\hat{\bar{\hvec}}_{q,m}$. Since this paper considers only \gls{zf}-type strategies, the factor $\E\{\sqrt{1-Z_{q,m}}\}$ does not affect the beamformer, and therefore, is omitted.}
  For  analytical simplicity, we make the following  assumption. 
\begin{assumption}
\label{ass_power_standard_shceme}
 Each \gls{srrh} transmits equal power to every \gls{ms}; i.e., \(P_q = P_{q,m}=P,
\forall q\in \cal{Q}\), \(m\in \cal{M}\) (cf. \cref{ass_power_constraint} for \(P_q\)).\footnote{Note that the equal power assumption is not optimal (see, e.g. \cite{Yu2019}) and is made to simplify the analysis, which is already very complicated.} To prevent \glspl{srrh} from violating  their  power constraint \(P_{\rm max}\),  we set   \(P=\frac{P_{\rm max}}{Q}\).
Moreover,   \(n_q \sim \calC \calN(0,1)\) (cf. \eqref{Signal-at-User-q}). 
\end{assumption}

From \eqref{Signal-at-User-q} and  \cref{ass_power_standard_shceme}, the \gls{sinr} at \gls{ms}-\(q\) is
\begin{IEEEeqnarray}{rCl}
        \label{eq_SignalModelq}
        {\rm  SINR}_q(\{\hat\pvec_{i}\}_{i\in {\cal Q}})\define\frac{P\vert \hvec^{\dagger}_q \hat\pvec_q \vert^2}{1 + P\sum_{j\in\mathcal{Q}_{\mbox{\tiny-}q}} \vert \hvec^{\dagger}_q \hat\pvec_j \vert^2},
        \;   \forall q\in{\mathcal{Q}}\;\;
\end{IEEEeqnarray}
where \(\hat\pvec_{q}\define[\hat\pvec_{q,1 }^{\dagger},...,\hat\pvec_{q,M}^{ \dagger}]^{\dagger}
\in\comp^{M N_t \times 1}\), and $\hat\pvec_{q,m}$ is the estimate of $\pvec_{q,m}$ under    \gls{rvq}, similar to \eqref{eq_ Expression for the estimated vector}.  Assuming a Gaussian codebook and  \gls{sud}, the achievable ergodic rate of \gls{ms}-\(q\) {with imperfect \gls{csi}} is
\begin{IEEEeqnarray}{rCl}
\label{eq_TrueRate}
 \hat R_q\define\E\big\{\log\big(1
+ {\rm  SINR} _q(\{\hat\pvec_{i}\}_{i\in
{\cal Q}})\big)\big\}\;\;
\end{IEEEeqnarray}
 The expectation is with respect to the joint distribution of the channel  and  the \gls{rvq}'s random codebook.\footnote{Note that \(\pvec_i\) is a function of the quantization \(\hat\hvec_{q}\), which, in the case of \gls{rvq}, is a function of the channels \(\{\hvec_{ q,m}\}_{m\in \cal{M}}\) and of the codebook generated for each channel. Moreover,  \(\hat\pvec_i\) is a function of \(\{\pvec_{ i,m}\}_{m\in \cal{M}}\).}
To evaluate performance, later we will compare   $\hat R_q$ to the corresponding throughput $R^{\star}_q$ without quantization error; i.e., 
\begin{IEEEeqnarray}{rCl}
        \label{rateWithPerfectP}
 R^{\star}_q \define  \E\big\{\log\big(1
+ {\rm  SINR} _q(\{\pvec^{\star}_{i}\}_{i\in
{\cal Q}})\big)\big\}\;\;
\end{IEEEeqnarray}
where  \(\pvec^{\star}_q\) is given in  \eqref{eq_ZFP}
while substituting 
$\hat \hvec_{q}=\hvec_{q},
\forall q\in{\cal Q}$, and  is assumed to be  fed-back perfectly. 
 \begin{theorem}
      Consider 
\cref{ass_channelModel,ass_long_term_channel,ass_quatnization,ass_quatnization_back,ass_power_standard_shceme}
and define the rate loss as
\begin{IEEEeqnarray}{rCl}
\label{rateGap} 
\Delta R_q \define R^{\star}_q-\hat R_q
\end{IEEEeqnarray} 
where $\hat R_q$ and $R^{\star}_q$ are defined in \eqref{eq_TrueRate} and \eqref{rateWithPerfectP}, respectively. Consider some $q\in {\cal Q}$ and   
        assume that \(\hvec_{q',m}\), \(\pvec_{q',m}\) are  quantized with \(B/Q\in\nat\) bits (cf.  \cref{ass_quatnization,ass_quatnization_back}),  each, \(\forall q'\in{\cal Q}, \forall m\in{\cal M}\);
        then
\begin{IEEEeqnarray}{rCl}
        \label{eq_Rate_Gap_Bound_StandardScheme}
\Delta
R_q\leq &&\Delta \bar{R}_{1,{q}}+\Delta\bar{R}_{2,{q}}
\end{IEEEeqnarray}
where 
\begin{IEEEeqnarray}{rclll}
        \label{eq_define_DeltaR_1}
        \Delta \bar{R}_{1,q}&\define&\log \Big\{1 +\frac{{  \alpha_{q}} PN_t(Q - 1)}{M}
        \times\Big[
        \frac{1}{N_t-1}
        \Dcond{\nonumber\\&&~\times}{}2^{\frac{-B}{Q(N_t-
                        1)}}\Big(2\big({ 1-\mathcal{U}(2^{B/Q},a) }\big)+ 2^{\frac{-B}{Q(N_t - 1)}}\Big)\nonumber\\&&~+\big(1-{\cal U}(2^{B/Q},a)\big)^{2}-\big(1-{\cal
                U}(2^{B/Q},a)/   2-{\cal
                U}(2^{B/Q},2a)\big)^{4}\Big]\Big\}
        \\
        \Delta \bar {R}_{2,q}&\define & 6 R_q^{\star} {\cal \big(U}(2^{B/Q},a)+{\cal V}_{M}(2^{B/Q},a)\big) 
\nonumber        \\&&+
\frac{\pi
                {\alpha_{q}} P N_{\rm t} }{ \sqrt{{\alpha_{q}}PN_{\rm t}+1}}\big(  {\cal U}(2^{B/Q},a/2)+{\cal V}_{M}(2^{B/Q},a/2)\big)      \label{eq_define_DeltaR_2}
\end{IEEEeqnarray}
Here ${  \alpha_{q} \define \sum_{m=1}^M \alpha_{q,m}}$, $a=\frac{1}{N_{t}-1}
$ and\footnote{Note that \(\E\{Z^i_{q,m}\} ={\cal U}(2^{B/Q},ai) \), where \(Z_{q,m}\) is defined in \eqref{eq_DecompositionJindalhBar}.}
 \begin{IEEEeqnarray}{rCl}
 \label{eq_Define_U}
{\cal U}(x,a)&=&x \beta(x,1+a)
 \\{\cal V}_{M}(x,a)&=&\frac{(M-1)}{\sqrt{2
M-1}} 
 \sqrt{{\cal U}\big(x,2 a\big)- {\cal U}\big(x,a\big)^2}
\label{eq_define_U2} 
\end{IEEEeqnarray} 
where $\beta(\cdot)$ is the Beta function.
 \label{Theorem:Theorem1}
\end{theorem}
\begin{remark}
 Under   \cref{ass_channelModel}, the perfect-\gls{csi} rate, $R^{\star}_{q}$, can be calculated based on known results.  For example, consider the case where the long-term channel-attenuation is equal for each \gls{srrh}; i.e., $\alpha_{q,m}=\alpha_{q,m'}$ $\forall m,m'\in {\cal M}$, and without loss of generality, assume  that $\alpha_{q,m}=1/M$ $\forall m\in{\cal M}$.  In this case, it is straightforward to show that
\begin{IEEEeqnarray}{rcl}
\label{eq_no_quanization_rate}
 R_q^{\star}= R^{\star}=\E\{\log\big(
1+P\vert\hvec^{\dagger}_q\pvec^{\star}_q
        \vert^2\big)\big\}=
(\log e)e^\frac{M}{P}\sum_{k=0}^{T
 -1}\Gamma
\Big(-k,\frac{M}{P}
\Big)\Big(\frac{M}{P}\Big)^{k}\define
\varphi(T,P/M) 
\end{IEEEeqnarray}
where, $T=MN_{t}-(Q-1)$ 
 and \(\Gamma\left(\cdot,\cdot\right)\)
is
the incomplete Gamma
function. In the case where $\exists ~ m\neq m'$ such that  $\alpha_{q,m}\neq\alpha_{q,m'}$,  an expression for $R^{\star}_{q}$  is complicated.   A closed-form expression can be found in \cite{Bas2013} (after straightforward adaptations to \gls{zf}) in the two-user case. For more than two users, such an expression is too complicated;  nevertheless, it  can be approximated, see  \cite{Basnayaka2012b} Sec. IV.A for the two-user case and \cite{Basnayaka2013,Senanayake2015a} for more than two users.  
\label{Comm_rate_perfect_csi}
\end{remark}

\begin{IEEEproof}[Proof of  \cref{Theorem:Theorem1}]
\label{ProofTheorem1}
 By the  assumptions of  \cref{Theorem:Theorem1} and using \eqref{eq_SignalModelq}, \eqref{eq_TrueRate},  \eqref{rateWithPerfectP}, and \eqref{rateGap}, it follows that
        \begin{IEEEeqnarray}{rcl}\label{eq_Delta_R_with_A1A2A3theorem_1_FirstRow}
                \Delta R_q
                &=& \E\left\{\log \left(1 + P \vert \hvec^{\dagger}_q \pvec_q^{\star} \vert^2 \right)\right\}\Dcond{ \nonumber\\ &&}{}- 
                \E\Big\{\log \big(1 + P \vert \hvec^{\dagger}_q \hat\pvec_q \vert^2 + P\sum_{j\in\mathcal{Q}_{\mbox{\tiny-}q}} \vert \hvec^{\dagger}_q \hat\pvec_j \vert^2 \big)\Big\}\nonumber\\
                &&+ \E \Big\{\log \Big(1 + P\sum_{j\in\mathcal{Q}_{\mbox{\tiny-}q}} \vert \hvec^{\dagger}_q \hat\pvec_j \vert^2 \Big)\Big\}\Dcond{\nonumber\\&
                        \le&}{\le} A_1- A_2
                + A_3
       \label{eq_Delta_R_with_A1A2A3theorem_1} \end{IEEEeqnarray}
        where \(\hat\pvec_{q}\) and $\pvec^{\star}_q $ are defined in \eqref{eq_SignalModelq}  
 and \eqref{rateWithPerfectP}, respectively, and
$
A_1 = \E\{\log
 (1 + P \vert \hvec^{\dagger}_q 
 \pvec^{\star}_q \vert^2 )\}$, 
  $
  A_2 = \E\big\{\log \big(1 + P\vert \hvec^{\dagger}_q \hat\pvec_q \vert^2 \big)\big\}$, 
$
A_3 = \E \{\log (1 + P\sum_{j\in\mathcal{Q}_{\mbox{\tiny-}q}} \vert \hvec^{\dagger}_q \hat\pvec_j \vert^2 )\}
$.
         The inequality \eqref{eq_Delta_R_with_A1A2A3theorem_1} follows because \(P\sum_{j\in\mathcal{Q}_{\mbox{\tiny-}q}} \vert \hvec^{\dagger}_q \hat\pvec_j \vert^2 \geq 0\) and \(\log(1+x)\) is a monotone increasing function.
The desired result \eqref{eq_Rate_Gap_Bound_StandardScheme} then follows from the following lemmas. 
\begin{lemma}\label{Lemma_Lemma2_BoundonC}
                Under  the assumptions of \cref{Theorem:Theorem1}, 
    $A_{3}\leq \Delta \bar R_{1,q}$ (cf. 
  \eqref{eq_Delta_R_with_A1A2A3theorem_1}, \eqref{eq_define_DeltaR_1}).
        \end{lemma}
        \begin{IEEEproof}
                See Appendix \ref{App:proof of
Lemma2:Bound
        on C}.
        \end{IEEEproof}
         \begin{lemma}
        \label{Lemma:Lemma3 Bound A-B}
                Under the assumptions of \cref{Theorem:Theorem1}, 
 $A_{1}-A_{2} \leq \Delta \bar R_{2,q}$ (cf. 
  \eqref{eq_Delta_R_with_A1A2A3theorem_1}, \eqref{eq_define_DeltaR_2}).
 \end{lemma}
 \begin{IEEEproof} See Appendix \ref{AppendixProofLemma3}.
        \end{IEEEproof}

Substituting \cref{Lemma_Lemma2_BoundonC,Lemma:Lemma3
Bound A-B}  into \eqref{eq_Delta_R_with_A1A2A3theorem_1} 
establishes \eqref{eq_Rate_Gap_Bound_StandardScheme}.
\end{IEEEproof}

%
%

{The following corollary simplifies the  rate-gap bound in \cref{Theorem:Theorem1} as the number of quantization bits gets large.}
\begin{corollary} 
The bound \(\Delta  R_q\leq \Delta \bar{R}_{1,{q}}+\Delta\bar{R}_{2,{q}}\) in \eqref{eq_Rate_Gap_Bound_StandardScheme} can be further approximated as 
\begin{IEEEeqnarray}{rCl}
\Delta \bar{R}_{1,{q}}+\Delta\bar{R}_{2,{q}}& =& 2^{\frac{-B}{2Q \left({N}_t-1
  \right)}}
  \frac{\pi{\alpha_{q}} P N_{t} }{ \sqrt{{\alpha_{q}}PN_{t}+1}}
 \left[V_{M}\Big({a}/{2}\Big)+\Gamma ( {a/2}+1) \right]+O(2^{\frac{-B}{Q \left({N}_t-1
  \right)}})
\label{eq_rough_approximation}
\end{IEEEeqnarray}
where $\Gamma(\cdot)$ is the Gamma function and $V_{M}(a)=\sqrt{\Gamma (2 a+1)-\Gamma
(a+1)^2}  (M-1)/\sqrt{2 M-1}$.
\label{corolary_closed_form_rough_approximation}
\end{corollary}
\begin{IEEEproof}  
Let $z=2^{B/Q}$ and denote
\begin{IEEEeqnarray}{rCl} \label{eq_Delta_R_z}
\Delta \bar{R}_{1,q}(z)+ \Delta\bar{R}_{2,q}(z)=\log \left(1+W_1(z)+W_2(z)\right) +W_3(z)
\end{IEEEeqnarray}
 where \(\Delta\bar{R}_{1,q}\) and \(\Delta\bar{R}_{2,q}\) are defined in \eqref{eq_define_DeltaR_1} and \eqref{eq_define_DeltaR_2}, respectively, and $W_1(z)=  {\alpha_{q}}P (Q-1) z^{-a} N_t \big(2 (1-\mathcal{U}(z,a))+z^{-a}\big)/{M
\left(N_t-1\right)},$ $W_2(z)={\alpha_{q}}P (Q-1)N_t/M\big[(1-\mathcal{U}(z,a))^2-\big(1- \mathcal{U}(z,a)/2-\mathcal{U}(z,2
a)\big)^4\big],$ $W_3(z)=\frac{\pi{\alpha_{q}} P N_{t} }{ \sqrt{{\alpha_{q}}PN_{t}+1}} \big[{\cal U}\big(z,a/2\big) + {\cal V}_M\big(z,a/2\big)\big]+ 6 R_q^{\star} \big[{\cal U} (z,a) + {\cal V}_M(z,a)\big]$.
It can be shown that 
\begin{IEEEeqnarray}{rCl}
\label{eq_asymptotic expression_for_U}
\mathcal{U}(z,a)=\Gamma (a+1) z^{-a}+O(z^{-a-\frac{1}{2}}).
\end{IEEEeqnarray} 
By substituting the latter into  $W_1(z)$, it can be shown that $W_1(z)=\frac{2{\alpha_{q}}   P (Q-1) z^{-a} N_t}{M \left(N_t-1\right)}+O(z^{-2
a})$.  
Now   to $W_2(z)$. Substituting, \eqref{eq_asymptotic expression_for_U}, it can be shown that $W_2(z)={\alpha_{q}}P (Q-1)N_t/M\big[(1-\Gamma (a+1) z^{-a})^2-(1- \Gamma (a+1) z^{-a}/2)^4\big].$ Moreover, because    $(1-x)^2-\left(1-{x}/{2}\right)^4=-\frac{x^4}{16}+\frac{x^3}{2}-\frac{x^2}{2}$, it follows that $W_2(z)=O(z^{-2 a}).$
 Finally, we turn to $W_3(z)$. Because  
${\cal U}(z,a) + {\cal V}_M(z,a)= \big(V_{M}(z,a)+\Gamma(a+1)\big)z^{-a}+O(z^{-a-1}),$
 it follows that $W_3(z)=\frac{\pi{\alpha_{q}} P N_{t} }{ \sqrt{{\alpha_{q}}PN_{t}+1}}z^{-\frac {a}{2}} \times\big[V_{M}\big(a/2\big)
 +\Gamma\big(1+a/2\big)\big] +6 R_q^{\star} z^{-a} \big[V_{M}(a)+\Gamma (a+1)\big]+O(z^{-1-\frac a 2}).$
Then, by substituting $W_{1}$, $W_{2}$ and $W_{3}$ into \eqref{eq_Delta_R_z} while taking lower order terms, the desired result follows.  
\end{IEEEproof}
  
We conclude this section with some insights. From \cref{corolary_closed_form_rough_approximation}, it follows that the rate-gap decreases at the rate $2^{\frac{-B}{2Q \left({N}_t-1
                \right)}}$ as $B$ increases. Furthermore, note that $\frac{\pi{\alpha_{q}} P N_{t} }{2 \sqrt{{\alpha_{q}}PN_{t}+1}}=O(\sqrt P)$ as $P\rightarrow\infty $. Therefore, to maintain
the overall number of degrees of freedom, $2^{\frac{-B}{2Q(N_{t}-1)}}$ should decrease at least like $\sqrt P$; i.e.,
the number of bits per channel should, at least, increase linearly with the
\gls{snr} in dB as well as with the number of \glspl{ms}. Otherwise, the
network is interference limited.
This result is consistent with previous findings on  the single-Tx broadcast
channel (cf. \cite{jindal2006mimo}). Finally, the rate gap decrease  $2^{\frac{-B}{2Q \left({N}_t-1
                \right)}}$ implies that it
is possible to reduce the rate gap without increasing $B$ by having a smaller $Q$, or having an effective
number of antennas less than $N_t$. The latter insight is the motivation for the \gls{paq} \gls{csi} sharing scheme presented in the following section.

However,  while $\Delta R_q$ is improved if $N_t$ or $Q$  decreases, $R^{\star}_q$ deteriorates due to a loss in antenna gain. This trade-off determines if the achievable rate, $\hat R_{q}$ (cf. \eqref{rateGap}), increases or decreases.
In the sequel, we show that $\hat R_q$ can be drastically improved under a good precoding strategy in most cases. Numerical results for the proposed bounds
are given in Sec. \ref{SecNumericalResults}.

\section{The precode and quantize CSI  sharing   Scheme}
\label{HierarchicalCoMPDescription}
 The \gls{paq} \gls{csi} sharing scheme aims  to reduce \gls{csi} overhead in the L2-link and the fronthaul  information rate. Each \gls{srrh}, say \gls{srrh}-\(m\),  applies a front-end precoding matrix \(\Amat_{m}\in \mathbb{C}^{N_t\times {\tilde N}_{t}},\) calculated according to its local \gls{csi} \(\{\hvec_{q,m}\}_{q\in\mathcal{Q}}\). This
creates  effective low-dimensional channels \(\tilde\hvec_{q,m}^{\dagger}=\hvec_{q,m}^{\dagger}\Amat_{m}\in \mathbb{C}^{1 \times {\tilde N}_{t}}, \  \forall q\in \mathcal{Q}\), with 
${\tilde N}_{t}<N_t$, that can be quantized more accurately than $\hvec_{q,m}$ \citep{Noam2019Two}.  We further denote the overall effective channel as \(\tilde \hvec_{q}\define[{\tilde\hvec}^{\dagger}_{q,1} ,...,
{\tilde\hvec}
^{\dagger}_{q,M}]^{\dagger} \in \comp^{M {\tilde N}_{t}\times 1}\).

\begin{definition}[\gls{ms} allocation policy]
\label{Def_ms_allocation_stragegy}
  To determine  $\Amat_m$, \gls{srrh}-\(m\) picks a subset of the  \glspl{ms}
  $\bar {\cal S}_m\subset {\cal Q}$, where $\vert \bar {\cal S}_m\vert=\bar Q$, according to the   policy   detailed next. Knowing  $\{\alpha_{q,m}\}_{q\in{\cal Q}}$ S-RRH-$m$, picks $\bar Q$ MSs
that have the most significant  attenuation; that is, $\bar {\cal S}_{m}$ includes \glspl{ms}  such that $\alpha_{q,m}\leq\alpha_{q',m},
\forall q\in \bar{\cal S}_m,$ $q'\in{\cal Q}\setminus \bar{\cal
S}_m$.    
\end{definition}

 Given $\bar {\cal S}_{m}$, 
$\Amat_m$ is set as the projection matrix into the null space of
the matrix whose columns are given by ${{\{{{\mathbf{h}}_{q,m}}\}}_{q\in
    \bar{\mathcal{S}}_m}}$; i.e.,
\beq\label{eq_defineAm}
{{\mathbf{A}}_{m}}=[{{\mathbf{u}}_{{1}}}\cdots {{\mathbf{u}}_{{\tilde N}_{t}}}],\;
\uvec_{i}\in {\mathbb C}^{N_t \times
  1}
\eeq
where  ${\tilde N}_{t}={N_t}- \bar  Q$ and $\{{{\mathbf{u}}_{{i}}}\}_{i=1}^{{\tilde N}_{t}}$ is an orthonormal basis for the orthogonal complement of \({\rm span}( {{\{{{\mathbf{h}}_{q,m}}\}}_{q\in \bar{\mathcal{S}}_m}})\). 
 Thus,   
\gls{srrh}-\(m\) now serves only $Q- \bar Q$ \glspl{ms},   denoted by \({\mathcal S}_m={\mathcal Q}\setminus \bar{\mathcal S}_m \subset \mathcal{Q}\).\footnote{Under this policy,  \glspl{ms} may
remain unserved; i.e., $q\in \bar {\cal S}_{m}, \forall m\in {\cal M}$. In this case, these \glspl{ms} can be reallocated
at the expense of \glspl{ms} that are served by the largest number
of  \glspl{srrh}.} From \eqref{eq_defineAm}, and 
because each \gls{srrh} has perfect local \gls{csi}, $\tilde\hvec_{q,m}=\0_{\tilde N_t}$, $\forall q\in \bar{\mathcal S}_m$. Thus, \gls{srrh}-$m$
  now sends the \gls{jpcu} only $Q-\bar Q$ channels 
${{\{{{\tilde{\hvec}}_{q,m}}\}}_{q\in
\mathcal{S}_m}}\), of lower dimension $\tilde N_{t}<N_{t}$,  which can be
quantized more accurately.  Denote the 
  estimate of $\tilde \hvec_{q,m}$ at the \gls{jpcu} by $\hat{\tilde \hvec}_{q,m}$ and
\beq\label{eq_define_hat_tile_h_q}\hat{\tilde\hvec}_q \define[\hat{\tilde\hvec}^{\dagger}_{q,1},...,\hat{\tilde\hvec}
^{\dagger}_{q,M}]^{\dagger} \in\comp^{ M {\tilde N}_{t}\times 1}.\eeq   Since the \gls{jpcu} knows $\bar {\cal S}_{m}$\footnote{ Because  the \gls{jpcu} knows $\{\alpha_{q,m}\}_{m\in {\cal M},q\in{\cal Q}}$
  (cf. \cref{ass_long_term_channel}), it can determine  $\bar {\cal S}_m$ by
applying the policy given in \cref{Def_ms_allocation_stragegy}, and therefore also knows \(\tilde N_t\).} it also knows that $\tilde\hvec_{q,m}
= \bm{0}_{\tilde N_t},$ $\forall{q\in \bar{\mathcal{S}}_m},  m\in\mathcal {M}\); hence it only estimates  $\{\tilde \hvec_{q,m} \}_{{q\in {\cal S}_m},{m\in {\cal M}}}$, whereas $\{\hat{\tilde\hvec}_{q,m}\}_{{q\in \bar {\cal S}_m},{m\in {\cal M}}}$ are set to zero; i.e.,    
$\hat {\tilde\hvec}_{q,m}=\bm{0}_{\tilde N_t},$ $\forall m\in {\cal M},q\in\bar{\cal S}_{m}$.
  Upon receiving the \gls{csi} from all \glspl{srrh}, \(\{\hat {\tilde \hvec}_{q}\}_{q\in\mathcal{Q}}\), the \gls{jpcu} computes 
\(\{\tilde \pvec_{q}\}_{q\in\mathcal{Q}}\), where 
\begin{IEEEeqnarray}{rCl}
\label{EqOverAllBeamformer}
{{\tilde\pvec}_{q}}\define[{\tilde\pvec}_{q,1}^ {\dagger},...,{\tilde\pvec}_{q,M}^{\dagger}]
^{\dagger}\in\comp^{M {\tilde N}_{t}\times 1}
\end{IEEEeqnarray}  is the overall beamformer designated for \gls{ms}-\(q\).\footnote{
Since each \gls{srrh} only serves a subset of the \glspl{ms}, full data sharing is unnecessary. Note that $\tilde \hvec_{q}$ satisfies $\tilde \hvec_{q}=\tilde \hvec_{q}\odot
(\vvec_q\otimes \onevec_{{\tilde N}_{t}})$, where \(\vvec_q\)  is an $M$-dimensional vector satisfying   $[\vvec_{q}]_{m}=1$ if \gls{srrh}-\(m\) serves
\gls{ms}-\(q\),   and \(0\) otherwise ({in the standard scheme
every \gls{srrh} serves every \gls{ms}, hence \(\vvec _q = \bm{1}_{M},\  \forall
q\in \mathcal{Q}\)).} Therefore, if $\tilde\pvec_{q}\neq \tilde\pvec_{q}\odot
(\vvec_q\otimes \onevec_{{\tilde N}_{t}})$, it follows that some \glspl{srrh}, which do not serve \gls{ms}-$q$, do transmit $s_{q}$.  Explicitly, if  $\tilde \hvec_{q,m}=\0_{\tilde N_t}$ and  $\tilde\pvec_{q,m}\neq \0_{\tilde N_t}$ for some $m\in{\cal M}$,  \gls{srrh}-$m$ must transmit the signal $s_{q}$, which \gls{ms}-$q$ does not receive.  To avoid  transmitting more data than 
necessary, we set the beamformer \({\tilde\pvec}_{q}\) orthogonal to \(\{\hat{\tilde\hvec}_j\odot
(\vvec_q\otimes \onevec_{{\tilde N}_{t}})\}_{j\in\mathcal{Q}_{\mbox{\tiny-}q} }\) from which it follows that  \({\tilde\pvec}_{q}
={\tilde\pvec}_{q}\odot (\vvec_q\otimes \onevec_{{\tilde N}_{t}})  
  \); i.e., the beamformer's weights corresponding to  \glspl{srrh} that 
do not serve \gls{ms}-\(q\) are   zero. By not  sending $\{s_{q}\}_{q\in \bar{\cal S}_{m}}$ to  \gls{srrh}-$m$,   we reduce the number of data streams for that \gls{srrh} to  $Q-\bar Q$, rather
than $Q$ as in the standard scheme.
\label{RemarkZeroChannels}
}

 \begin{definition}
  \label{def_ZF-P-P-and-Q}
The \gls{paq} beamformer for \gls{ms}-$q$ is $
 \tilde\pvec_q \define  \tilde\Nmat_q \frac{(\hat{\tilde\hvec}^{\dagger}_q \tilde\Nmat_q)^{\dagger}}{\|\hat{\tilde\hvec}^{\dagger}_q\tilde\Nmat_q\|}
, \forall q\in\mathcal{Q}$, where    
\(\tilde\Nmat_q \in \comp^{M \tilde N_{t} \times M {\tilde N}_{t} -( \tilde Q_q-1)}\) is the projection matrix into the null space of \(\{\hat{\tilde\hvec}_j\odot (\vvec_q\otimes \onevec_{{\tilde N}_{t}})
 \}_{j\in\mathcal{Q}_{\mbox{\tiny-}q} }\). The factor $\tilde Q_{q}$
is the number of \glspl{ms} such that $\tilde \hvec_{q}^{\dagger}\tilde
\hvec_{j}\neq0$, $\forall q, j\in {\cal Q}$; i.e.,       $\tilde
Q_q=Q -\sum_{j\in{\cal
                        Q}_{\mbox{\tiny-}q}} \chi_{\{0\}}(M_{q,j})$,
where   \(M_{q,j}\) is the number  of \glspl{srrh} that serve
both \gls{ms}-\(q\) and \gls{ms}-\(j\).\label{Def_P_Q_beamformer}\footnote{The
 coefficient  $\tilde Q_{q}$ (cf. \cref{def_ZF-P-P-and-Q})
is  the number of \glspl{ms} served by at least one
of the \glspl{srrh} that serve \gls{ms}-$q.$ $\tilde Q_{q}-1$ is the number of \glspl{ms} to which the ZF precoder must zero the interference inflicted by \gls{ms}-\(q\).
}
\end{definition}
 After setting $\tilde\pvec_{q}$ according to \cref{def_ZF-P-P-and-Q}, the \gls{jpcu} quantizes \({\tilde\pvec}_{q,m}\), (cf. \eqref{EqOverAllBeamformer}) and sends each \gls{srrh} its relevant precoders. Moreover, because
$\{\tilde \pvec_{q,m} \}_{{q\in \bar S_m},{m\in {\cal M}}}=\bm{0}_{\tilde N_t},$ the \gls{jpcu}
does not have to send  \gls{srrh}-$m$ the entire set  $\{{\tilde\pvec}_{q,m}\}_{q\in {\cal
Q}}$, but rather  sends    
${{\{{{\tilde{\pvec}}_{q,m}}\}}_{q\in
\mathcal{S}_m}}\), which consists  solely of  $Q-\bar Q$ beamformers. In more explicit terms, it  sends the quantization of $\{{\tilde\pvec}_{q,m}\}_{q\in {\cal  S }_m}$ to \gls{srrh}-$m.$ Since the latter have a lower dimension $\tilde N_{t}<N_{t}$, 
they  can be
quantized more accurately. Once having received these quantizations, \gls{srrh}-$m$ sets its overall beamformer toward 
\gls{ms}-\(q\) as
\begin{IEEEeqnarray}{rCl}
        \label{Eq_Overall_Beamformer_P_Q}
        \hat\pvec^{\rm\gls{paq}}_{q,m}
        \define \Amat_m \hat{\tilde\pvec}_{q,m} \in \comp^{N_t\times 1}
\end{IEEEeqnarray}
 where $\hat{\tilde\pvec}_{q,m}$
denotes the estimate of $\tilde\pvec_{q,m}$.  
\begin{definition}
  \label{def_overall-ZF-P-P-and-Q}
  The overall \gls{paq} beamformer $\hat\pvec_q^{\rm\gls{paq}}  
\in \comp^{MN_{t}}
\times 1$ for \gls{ms}-$q$  is  
\(\hat\pvec_q^{\rm\gls{paq}}\define\Amat\hat{\tilde\pvec}_{q},\) where  \(\Amat = {\rm blockdiag}\{\Amat_1,\Amat_2,...,\Amat_M\}
\in \comp^{MN_{t} \times M \tilde N_{t}}\) and     $\hat{\tilde \pvec}_{q}\define[\hat{\tilde \pvec}^{\dagger}_{q,1},\ldots,\hat{\tilde\pvec}^{\dagger}_{q,M}]^{\dagger}\in\comp^{M {\tilde N}_{t}\times 1}$.
\end{definition}
Note that because \({\tilde\pvec}_{q}
={\tilde\pvec}_{q}\odot (\vvec_q\otimes \onevec_{{\tilde N}_{t}})  
  \) it follows that \({\hat{\tilde\pvec}}_{q}= {\hat{\tilde \pvec}}_{q}\odot (\vvec_q\otimes \onevec_{{\tilde N}_{t}})\).
\begin{definition}
    \label{definitions_of_m_q_and_m_q_j}
    Let ${\cal M}_{q}$ be the set of \glspl{srrh}
that serve \gls{ms}-$q$; that is, ${\cal M}_{q}=\{m\in{\cal M}: q\in{\cal S}_{m}\}$ and  denote $M_{q}=\vert{\cal M}_{q}\vert$. Furthermore, let ${\cal M}_{q,j}\define {\cal
M}_{q}\cap{\cal M}_{j}$ be the set of \glspl{srrh} that serve both \gls{ms}-\(q\) and \gls{ms}-\(j\),  and denote  \(M_{q,j}=\vert{\cal M}_{q,j}\vert\).
\end{definition}

Substituting \(\hat\pvec_i^{\rm\gls{paq}}\) for \(\pvec_i, i \in{\cal Q}\) in \eqref{Signal-at-User-q},  \gls{ms}-\(q\) observes the signal
\begin{IEEEeqnarray}{rCl}
\label{eqSignalModelPaQ}
y_{q}= \tilde \hvec^{\dagger}_{q}\hat{\tilde \pvec}_{q}s_{q} +
\sum_{j\in\mathcal{Q}_{\mbox{\tiny-}q}}\tilde \hvec^{\dagger}_{q}\hat{\tilde \pvec}_{j}s_{j} + n
_{q}
\;, \quad\forall q\in{\mathcal{Q}}
\end{IEEEeqnarray}
where \(\tilde \hvec_{q}=[{\tilde\hvec}^{\dagger}_{q,1} ,...,
{\tilde\hvec}^{\dagger}_{q,M}]^{\dagger} \)
and \(\hat {\tilde \pvec}_{q} \) is given in \cref{def_overall-ZF-P-P-and-Q}.  {We note that  $\tilde\hvec_{q}$ replaces $\hvec_{q}$ because each \gls{srrh} applies $\Amat_{m}$  (cf. \eqref{eq_defineAm}); moreover,  the sum runs over $Q_{\text{-}q}$ because of the particular choice of $\Amat_{m}$  and $\tilde\pvec_{j}, j\in{\cal Q}$ (\cref{def_ZF-P-P-and-Q}), as discussed in \cref{RemarkZeroChannels}.}
The latter can be written as    
        $y_q =\sum _{m\in
\mathcal{M}_q} \tilde\hvec_{q,m}^{\dagger
}\hat{\tilde\pvec}_{q,m} s_q{}+\sum_{j\in\mathcal{Q}_{\mbox{\tiny-}q}}\sum _{m\in{\cal M}_{q,j}} \tilde\hvec_{q,m}^{\dagger }\hat{\tilde\pvec}_{j,m} s_j+n_{q}
,$ where  ${\cal M}_{q}$ and ${\cal M}_{q,j}$ are given  in \cref{definitions_of_m_q_and_m_q_j}.

The advantage of the proposed scheme is twofold. From \cite{jindal2006mimo}, it is known that when quantizing an   $N$-dimensional uncorrelated Rayleigh fading channel with $b$ bits, the quantization error is bounded above  by $2^{-\frac b{N-1}}$. Therefore,  the \gls{paq} has a smaller \gls{csi}-quantization error because the channels and beamformers  are ${\tilde N}_{t}$-dimensional, rather than ${N_t}$. Furthermore, since  each \gls{srrh} serves fewer \glspl{ms}, fewer channels and beamformers are delivered to the \gls{jpcu} and \glspl{srrh}, respectively, through the limited-rate links. Considering  an overall budget of \(B\) bits for each \gls{srrh}, it follows that the \gls{paq} scheme allocates each channel \(B/(Q - \bar Q)\) bits rather than $B/Q$  in the standard scheme. Consequently, the quantization error is bounded by $2^{-\frac{B}{(Q-\bar Q)(\tilde N_{t}-1)}}$ rather than by $2^{-\frac{B}{Q(N_{t}-1)}}$. The second advantage of the \gls{paq} scheme is in reducing fronthaul data load, which is a major problem in   \gls{cran}. This reduction is because
each \gls{srrh} serves only  \(Q-\bar Q\) \glspl{ms}. Hence, fewer data signals must be transferred via the fronthaul between the \gls{bbu} to each \gls{srrh}. Moreover, because each \gls{srrh} now serves fewer \glspl{ms}, the overall power allocated for each \gls{ms} may be increased.

\section{The \gls{paq} scheme: performance analysis}
 \label{SecPandQPanalysis}
We now present  results  corresponding to  \cref{Theorem:Theorem1} and \cref{corolary_closed_form_rough_approximation} for the  \gls{paq} scheme. We assume the  following. 
\begin{assumption}
Each \gls{srrh} serves  \(Q -\bar
Q\) \glspl{ms}   with  
\(\tilde P_{q,m}=\tilde P=\frac{ P_{\rm max}}{Q -\bar
Q }, \forall q\in \cal{Q}\), \( m\in \cal{M}$.
\label{ass_p&q_power}\end{assumption}

Because the analysis of 
the \gls{paq} is more complicated than the standard scheme, we  simplify the setup as follows. 
\begin{assumption}[symmetric system-geometry with an equal pathloss constrain]
 The long-term channel attenuation  satisfies
$\alpha_{q,m} = 1/M, \forall q\in {\cal Q},\; m\in{\cal M}$.
\label{ass_equal_alphas}\end{assumption}
  This assumption holds, e.g., if one places \glspl{srrh}  on the edges of a regular
polygon with $M$ nodes and
\glspl{ms} 
close to each other at the
center of that polygon.\footnote{In
more explicit terms, \glspl{srrh}
are placed at ${\cal G}=\{(r
\cos(2\pi m/M),r\sin(2\pi
m/M))\in\real^{2}: m\in\{0
,...,M-1\}\}$, where $r>0$
is fixed where
$(0,0)$ is the center of the
polygon. The \glspl{ms} are
placed very close to each
other around the point $(0,0)$.} 
Then, \glspl{ms} have  approximately the same long-term channel attenuation to each \gls{srrh}.  In a rich scattering
environment, the \glspl{ms} will experience independent fading.\footnote{This
happens as long as the distance
between them is larger than
the wavelength; which is a very reasonable since cellular wavelengths
are typically on the order of  centimeters.}  
 \begin{definition}\label{Def_P_and_Q_Rate}
 Consider Assumptions \ref{ass_p&q_power} and \ref{ass_equal_alphas}, let    $\Delta
\tilde R_{q} \define R_{q}^{\star}-\hat{\tilde
R}_{q}$   be  the \gls{paq} rate-gap where \(R^{\star}_{q}\) is given
in \eqref{rateWithPerfectP},
\(\hat{\tilde R}_{q} = \E\{\log(1+ {\rm SINR}_q(\{\hat{\pvec}_{i }^{\rm\gls{paq}} \}_{i\in {\cal Q}}))\}
  \)
     , and ${\rm SINR}_q(\cdot)$ 
is defined similarly to  \eqref{eq_SignalModelq}
with \(\hat\pvec_i^{\rm\gls{paq}}\) as its argument (cf. \cref{def_overall-ZF-P-P-and-Q}) while substituting  \(\tilde P\) for \( P\). 
\end{definition}
\begin{definition}
Let
$\tilde\pvec_q ^{\star}$ be
the \gls{paq} beamformer without
quantization error;
 i.e.,
$\tilde\pvec_q ^{\star}$ is
obtained by replacing $\hat{\tilde\hvec}_q$
 with $\tilde\hvec_{q}$ in  $\tilde\pvec_{q}$ (cf. 
 \cref{def_ZF-P-P-and-Q}) as well as in the calculation of $\tilde
\Nmat_{q}$.
We further denote the   \gls{paq}  inherent rate-loss by   $\Delta R_{{\rm AG},q} \define R^{\star}_{q}-\tilde R^{\star}_{q} $ where $R^{\star}_q$ is given in \eqref{rateWithPerfectP} and \(\tilde R^{\star}_{q} = \E\{\log
(1 + \tilde P \vert \tilde\hvec^{\dagger}_q
\tilde\pvec^{\star}_q \vert^2
)\}\). In other words, $\Delta R_{{\rm AG},q}$ is the difference between the  standard-scheme  and     \gls{paq}-scheme rates without quantization errors,  resulting from the loss in array gain.
\label{def_rate_loss_array_gain}
\end{definition}
 \begin{theorem}
\label{Theore:P and Q}
 Consider 
\cref{ass_channelModel,ass_long_term_channel,ass_quatnization,ass_quatnization_back,ass_p&q_power,ass_equal_alphas}, and assume that the     \gls{paq}
(cf. \cref{Def_P_Q_beamformer}) is  applied with   $B$ bits, (cf. \cref{ass_quatnization,ass_quatnization_back}),  where  \({\tilde \hvec}_{ q',m}\),  ${\tilde \pvec}_{ q',m}$
 are quantized using \gls{rvq} with    \(B/(Q -
\bar Q)\) (assumed integer) bits \(\forall q'\in{\cal Q}, \forall m\in{\cal M}\).
Then, the \gls{ms}-$q,\; q\in{\cal Q}$ expected
throughput-loss due to
\gls{csi} quantization,  in comparison  to a perfect \gls{csi}, satisfies
\begin{IEEEeqnarray}{rCl}
\label{InteBoundStandardScheme}
\Delta {\tilde R}_q &\leq& \Delta\bar  {\tilde{R}}_{1,q}+\Delta \bar{ \tilde R}_{2,q}+ \Delta R_{{\rm AG},q} 
\end{IEEEeqnarray}
where 
\begin{IEEEeqnarray}{rCl}
\Delta\bar  {\tilde{R}}_{1,q}&\define&\log\Big\{1 +\tilde P\Big[\sum_{j\in\Qmq}\frac{\tilde N_{t}{M_{q,j}}}{M_jM}\Big]
\times\Big[
        \frac{1}{\tilde N_{t}
        - 1}
        \Dcond{\nonumber\\&&~\times}{}2^{\frac{-B}{(Q-\bar Q)(\tilde N_t-
                        1)}}\Big(2\big({ 1-\mathcal{U}(2^{\frac{B}{Q-\bar Q}},\tilde a) }\big)+ 2^{\frac{-B}{{(Q-\bar Q)}(\tilde N_t - 1)}}\Big)\nonumber\\&&+\big(1-{\cal U}(2^{\frac{B}{Q-\bar
Q}},\tilde a)\big)^{2}-\big(1-{\cal
                U}(2^{\frac{B}{Q-\bar
Q}},\tilde a)/2-{\cal
                U}(2^{\frac{B}{Q-\bar Q}}
,2\tilde a)\big)^{4}\Big]\Big\}
\label{eq_define_DeltaR_1PandQ}
\\
\Delta \bar{\tilde R}_{2,q}&\define& 6 \tilde R^{\star}_q \big({\cal U}(2^{\frac
B{(Q-\bar Q)}},\tilde a)+{\cal
V}_{M_{q}}(2^{\frac {B}{Q-\bar Q}},\tilde a)\big) 
\nonumber        \\&&
+\frac{\pi \tilde P \tilde N_{t} }{ \sqrt{\tilde P \tilde N_{t}+1}}
\big({\cal U}
(2^{\frac B {Q-\bar Q}},{\tilde a}/{2})+{\cal V}_{M_{q}}(2^{\frac
B{Q-\bar Q}},\tilde a/2)\big)
     \label{eq_define_DeltaR_2PandQ}
\end{IEEEeqnarray}
Here,  $\tilde a=\frac 1{\tilde N_{t}-1}$, $\tilde P=\frac{P_{\max}}{Q -
 \bar Q}$, $M_{q}, M_{q,j}$ are in \cref{definitions_of_m_q_and_m_q_j} and ${\cal U}(\cdot)$, ${\cal
V}_{M}(\cdot)$ are defined in \cref{Theorem:Theorem1}.
 Furthermore, the term $\Delta R_{{\rm AG},q}$ (cf.  \eqref{InteBoundStandardScheme}  as well as \cref{def_rate_loss_array_gain}) satisfies  
\begin{IEEEeqnarray}{rCl}
\label{eq_AG_PandQ}
         \Delta R_{{\rm AG},q}= \varphi(T,P/M)- \varphi(      \tilde T_{q},\tilde P/M)
\end{IEEEeqnarray}
 where $\varphi(\cdot)$ and $T$ are  in \eqref{eq_no_quanization_rate}, 
$\label{eq_M_tilde_q}
\tilde T_{q}=M_{q}{\tilde N}_{t}-(\tilde Q_{q}-1)$,  and $\tilde
Q_{q}$  is  in \cref{def_ZF-P-P-and-Q}.

\end{theorem}
\begin{IEEEproof} 
         Similar to \eqref{eq_Delta_R_with_A1A2A3theorem_1}, it can be shown that 
\begin{IEEEeqnarray}{rCl}\label{eq_Delta_R_with_A1A2A3theorem_1_P_and_Q}
\Delta&& \tilde R_{q}\leq
\tilde A_{1}+ \Delta R_{{\rm AG},q}
- \tilde A_{2}+ \tilde A_{3}
\end{IEEEeqnarray}
where 
$\tilde A_{1}= \E\big\{\log\big(1 + \tilde P \vert \tilde\hvec^{\dagger}_q {\tilde{\pvec}^{\star}}_q \vert^2
\big)\big\}$,  
  $\tilde A_2 = \E\big\{\log \big(1 + \tilde P \vert \tilde\hvec^{\dagger}_q \hat{\tilde{\pvec}}_q \vert^2 \big)\big\}$,
$\tilde A_3 = \E \big\{\log \big(1 + \tilde P \sum_{j\in\mathcal{Q}_{\mbox{\tiny-}q}} \vert \tilde\hvec^{\dagger}_q \hat{\tilde{\pvec}}_j
\vert^2 \big)\big\}$ and the additional term is given by $\Delta R_{{\rm AG},q}=A_{1}-\tilde A_{1}$ where  \(A_1\) is defined in \eqref{eq_Delta_R_with_A1A2A3theorem_1}.            The proof then follows from the following lemmas.
\begin{lemma}\label{Lemma_Bound_on_A3_P_a_Q}
Under  the assumptions of \cref{Theore:P and Q}, $\tilde A_3\leq
\Delta \bar {\tilde R}_{1,q}$, and
 $                       \tilde A_{1}-\tilde A_{2}
\leq  \Delta \bar{ \tilde R}_{2,q}$ (cf. 
  \eqref{eq_define_DeltaR_1PandQ}, \eqref{eq_define_DeltaR_2PandQ}, \eqref{eq_Delta_R_with_A1A2A3theorem_1_P_and_Q}).
        \end{lemma}
        \begin{IEEEproof}
                See Appendix \ref{App_C_ProofOLemma_7_Bound_A3_P_Q}.
        \end{IEEEproof}

 After substituting the inequalities
of \cref{Lemma_Bound_on_A3_P_a_Q}  into \eqref{eq_Delta_R_with_A1A2A3theorem_1_P_and_Q} it remains to show \eqref{eq_AG_PandQ}. To this end,  we use $R^{\star}=\varphi(T,P/M)$ (cf. \eqref{eq_no_quanization_rate}) and     $   \tilde A_{1}=\varphi( \tilde T_{q},\tilde P/M)$,  obtained by applying the former while replacing $T$ and $P$ with $\tilde T_{q}$ and $\tilde P$, respectively.
 \end{IEEEproof}
\begin{definition}[Symmetric selection policy]\label{def_symm_policy} For $Q/M\in\nat$ and $M\bar 
  Q/Q\in\nat$, let  $\{{\cal
Q}_{m}\}_{m=1}^{M}$ be a  partition of ${\cal Q}$,
such that $\vert {\cal Q}_{m}\vert= Q/M\;\forall m$. In this selection 
policy,  $\forall m\in {\cal M}$, the  set of \glspl{ms} discarded by \gls{srrh}-$m,$ is    $\bar {\cal S}_{m}=\bigcup_{i=1}^{\bar 
QM/Q}{\cal Q}_{(i+m)\bmod M}$. \end{definition}
   \begin{corollary}
 \label{cor_P_Q_symmetric}
  Consider the assumptions of \cref{Theore:P and Q} and assume in addition that $Q/M\in \nat$, $M\bar 
  Q/Q\in\nat$ and a symmetric selection-policy (cf. \cref{def_symm_policy}).   Then 
$\Delta {\tilde R} \leq \Delta\bar  {\tilde{R}}_{1}+\Delta \bar{\tilde R}_{2}+ \Delta R_{{\rm AG}},$ where  $\Delta R_{{\rm AG}}=\varphi(
 T, P/M)-\varphi(
\tilde T,\tilde P/M),$  $\tilde{T}=M
(1-\bar Q/Q)
\tilde{N}_t+1-\min \big(Q,(2-{1}/{M}) Q-2 \bar{Q}\big)$, and
\begin{IEEEeqnarray}{rCl}
\Delta\bar  {\tilde{R}}_{1}&=&\log \bigg\{1 +\frac {\tilde P\tilde N_{t}(Q -\bar Q-1)}{M}\times\Big[
        \frac{1}{\tilde N_{t}
        - 1}
        \Dcond{\nonumber\\&&~\times}{}2^{\frac{-B}{(Q-\bar Q)(\tilde N_t-
                        1)}}\Big(2\big({ 1-\mathcal{U}(2^{\frac{B}{Q-\bar
Q}},\tilde a) }\big)+ 2^{\frac{-B}{{(Q-\bar Q)}(\tilde N_t - 1)}}\Big)\nonumber\\&&~+\big(1-{\cal
U}(2^{\frac{B}{Q-\bar
Q}},\tilde a)\big)^{2}-\big(1-{\cal
                U}(2^{\frac{B}{Q-\bar
Q}},\tilde a)/2-{\cal
                U}(2^{\frac{B}{Q-\bar Q}}
,2\tilde a)\big)^{4}\Big]\bigg\}
\label{DR1Symmetric}\\
\Delta \bar {\tilde R}_{2}&=&6 \varphi(\tilde T,\tilde
P/M)\big[ {\cal U}(2^{\frac
B{Q-\bar Q}},\tilde a)+{\cal
V}_{(1-\bar Q/Q)M}(2^{\frac {B}{Q-\bar Q}},\tilde a)\big]
\\&& \nonumber+
\frac{\pi \tilde P \tilde N_{t} }{ \sqrt{\tilde P \tilde N_{t}+1}}
\big[{\cal U}(2^{\frac B {Q-\bar Q}},{\tilde a}/{2})+{\cal V}_{(1-\bar Q/Q)M}(2^{\frac
B{Q-\bar Q}},\tilde a/2)\big]
     \label{eq_define_DeltaR_2PandQ_simple_case}
\end{IEEEeqnarray}
 \label{cor_P_and_Q_symmetric}
 \end{corollary}
\begin{IEEEproof}
Due to space limitations, we provide here an outline of the proof (a detailed proof is given in \cite{arad2020cran} Supplementary B). The first step shows that the sum in \eqref{eq_define_DeltaR_1PandQ} runs over constant terms, and can therefore  be replaced by a factor $Q-\bar Q-1$ in \eqref{DR1Symmetric}. To this end, one must show that $M_{q}=(1-\bar Q/Q)M,
\forall q\in{\cal Q}$. Finally, we substitute the latter result for $\tilde T_{q}$ in \eqref{eq_AG_PandQ} and obtain $\tilde T_{q}=\tilde T$, where $\tilde T$ is given in this corollary. 
\end{IEEEproof} 
\begin{remark}
Under the suppositions of \cref{cor_P_Q_symmetric}, all \gls{ms} rates are equal,  $\hat{\tilde R}_{q}=\hat{\tilde R}$ (cf. \cref{Def_P_and_Q_Rate}),  and satisfy $\hat {\tilde R} \geq R^{\star}
 -\Delta\bar  {\tilde{R}}_{1}-\Delta \bar {\tilde R}_{2}-
\Delta R_{\rm AG}$,
 where $R^{\star}$ is given in \eqref{eq_no_quanization_rate}.
Moreover, considering the standard  scheme  under  \cref{ass_equal_alphas} and $Q/M\in \nat$, it follows that 
$\hat { R} \geq R^{\star}
 -\Delta\bar  {{R}}_{1}-\Delta \bar R_{2}$.
\label{remark_4}
\end{remark}
 Next, similar to Corollary \ref{corolary_closed_form_rough_approximation}, we
have have following corollary.
\begin{corollary}
Consider $\Delta\bar  {\tilde{R}}_{1}$ and $\Delta \bar{\tilde R}_{2}$, given in \cref{cor_P_and_Q_symmetric}. Then the bound \(\tilde \Delta  R\leq \Delta \bar{\tilde R}_{1}+\Delta\bar{\tilde R}_{2}+ \Delta R_{{\rm AG}}\) in \eqref{InteBoundStandardScheme} can be further approximated as 
%
\begin{IEEEeqnarray}{rCl} 
 \Delta\bar  {\tilde{R}}_{1}+\Delta \bar{\tilde R}_{2}=&&2^{\frac{-B}{2(Q-\bar Q)
\left(\tilde {N}_t-1 
  \right)}}
 \frac{\pi \tilde P \tilde N_{t} }{ \sqrt{\tilde P \tilde N_{t}+1}}
\left[V_{M}\Big(\frac {1/2}{\tilde N_{t}-1}\Big)+\Gamma
\Big(\frac {1/2}{\tilde N_{t}-1}+1\Big)\right]+O(2^{\frac{-2B}{(Q-\bar Q) \left(\tilde {N}_t-1 
  \right)}})\;.
\label{eq_rough_approximation_paq}
 \end{IEEEeqnarray} 
\label{corolary_closed_form_rough_approximation_paq}
\end{corollary}
\begin{IEEEproof}The proof is identical to the proof of \cref{corolary_closed_form_rough_approximation}.
\end{IEEEproof}

We conclude this section with  a discussion and insights. By examining  Corollaries \ref{corolary_closed_form_rough_approximation}
and \ref{corolary_closed_form_rough_approximation_paq},
it follows that the rate loss $\Delta R$ in the standard scheme (which here is not a function of \(q\), cf. \cref{remark_4}) approaches zero as $B$ increases. . In contrast,
the rate gap in the \gls{paq} scheme,
$\Delta {\tilde R}$, is bounded
away from zero. Explicitly, it  approaches $ \Delta R_{{\rm AG}}>0 $ (cf. 
\cref{def_rate_loss_array_gain}), which is independent of $B$ and is due to the array-gain loss induced by the dimension reduction. However, the other terms $\Delta\bar 
{\tilde{R}}_{1}+\Delta \bar{\tilde R}_{2}$, comprising $\Delta \tilde R$, decrease to zero much faster 
than $\Delta\bar  {{R}}_{1}+\Delta \bar R_{2}$ (cf. \eqref{eq_rough_approximation} and
\eqref{eq_rough_approximation_paq}); therefore,  $\Delta \tilde R$ approaches $\Delta R_{{\rm AG}}$ much faster than $\Delta R$ approaches zero.
Subsequently,\   $\hat {\tilde R}$ approaches $R^{\star}-\Delta R_{\rm AG }$, much faster than $\hat R$ approaches $R^{\star}$. The final observation is that $\hat{\tilde R }$ can be higher than $\hat R$ as long as $\Delta R$ is more significant than $\Delta R_{\rm AG}$. Numerical results presented in the subsequent section indicate that $\hat {\tilde R}$ is indeed higher than $\hat R$ for a wide range of quantization bits.

\section{Numerical Results}
\label{SecNumericalResults} 

\begin{figure}
        \centering
\includegraphics[width=16 cm]{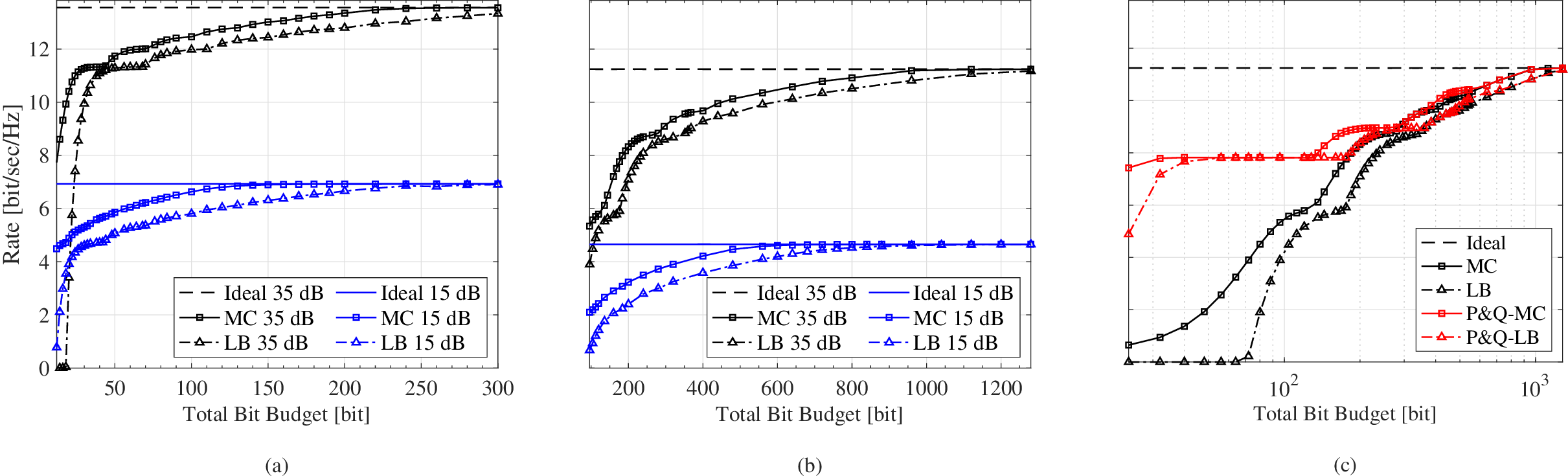}
        \caption{Averaged throughput lower bounds   (cf.  \cref{remark_4}) and \gls{mc} simulation, as a function of the overall bit budget \(B\) for \(N_t = 8\) antennas and \(M=4\) \glspl{srrh}, and equal
transmit-power to each \gls{ms}. Fig (a) considers \(Q=2\) \glspl{ms}  with non-equal pathlosses to each \gls{srrh},  (b) presents the symmetric configuration (cf. \cref{ass_equal_alphas})
                of \(Q = 8\) \glspl{ms}. In (c),  we compare the \gls{paq} (red)
to the standard scheme (black) for the same setup as (b) with \(P=35\) dB; hence the black curve is the same as in (b) but in a logarithmic scale. The vertical axis is the same in all figures.}
        \label{FigTheoreticSetupSimulation}
\end{figure}

\begin{figure}
        \centering
        \subfigure[\label{FigSystemSimulationA} Throughput vs.  $P_{\rm max}$ for  $B=176$ bits. ]
          { \includegraphics[width=\Dcond{0.4}{0.45}\textwidth]{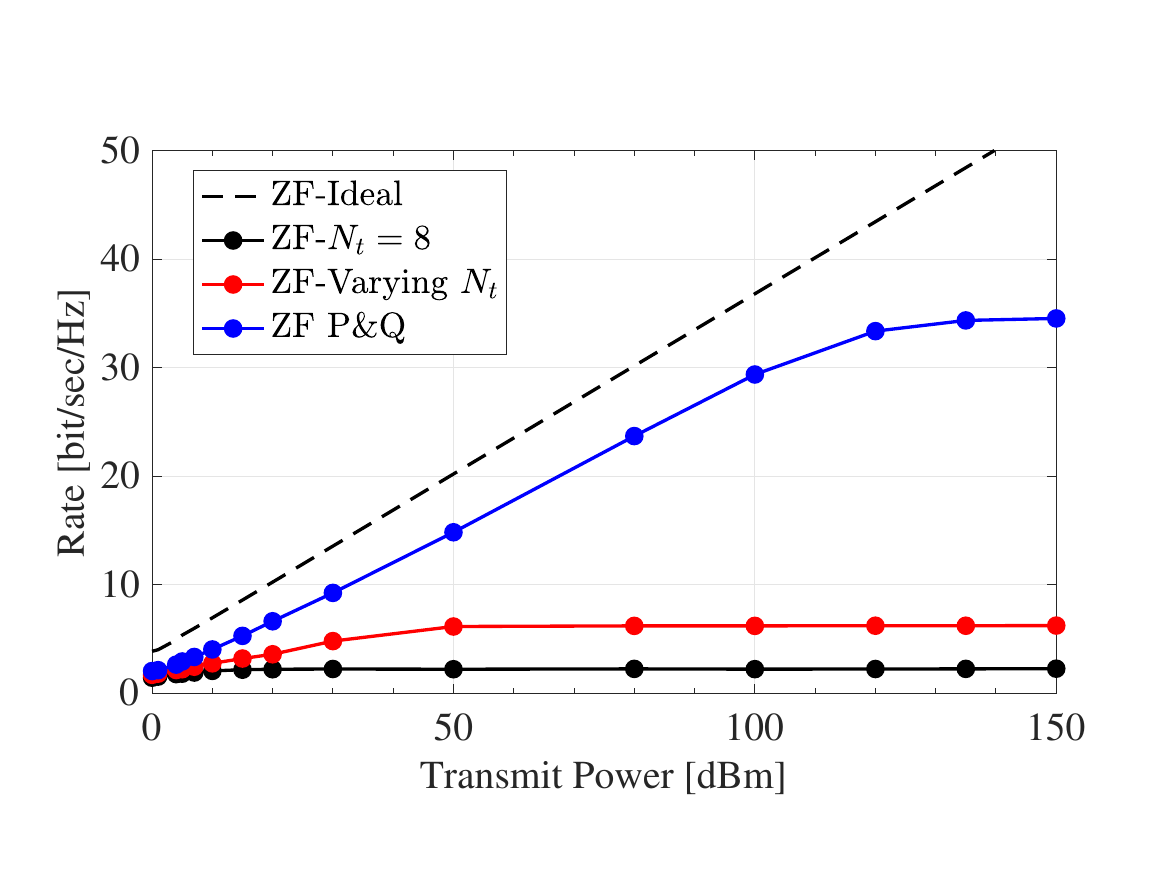}}
        \qquad
        \subfigure[\label{FigSystemSimulationB} Throughput vs.  $B$ for  $P_{\rm max}=37$  dBm. ]{\includegraphics[width=\Dcond{0.4}{0.45}\textwidth]{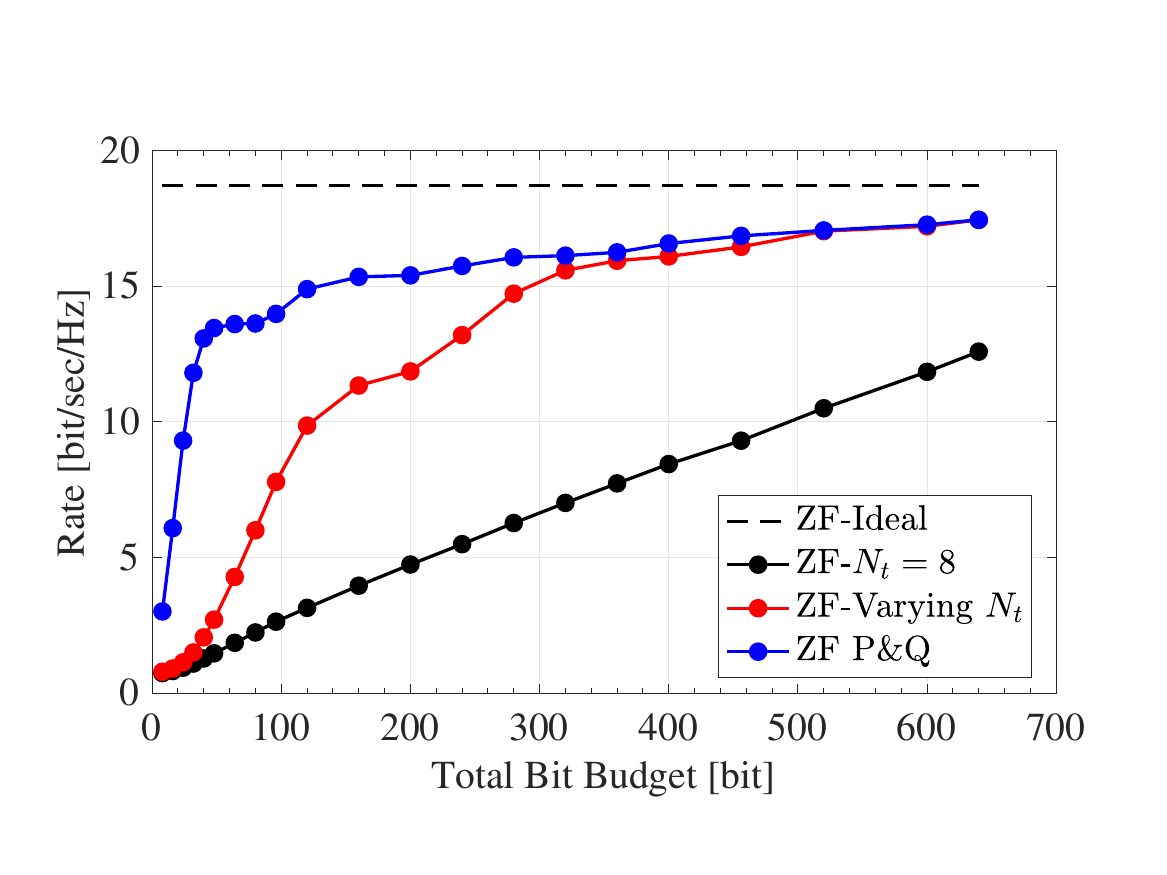}}
        \caption{ Throughput  of the \gls{paq} scheme (blue), standard scheme with varying $N_{{t}}$ (red), with fixed $N_{t}=8$ (black), and under perfect \gls{csi} (dashed).  We equally allocated the overall bit budget  
per \gls{srrh},  $B$ (cf. \cref{ass_quatnization,ass_quatnization_back})
to each \gls{ms}; that is, $B/Q$ and $B/(Q-\bar Q)$
 bits per \gls{ms} in the standard and in the \gls{paq} scheme, respectively, 
where non-integer
values  were floored.  }
        \label{FigSystemSimulation}
\end{figure}

In this section, we study two setups. The first consider setups matching the theoretical analysis in \cref{SecAnalysisResidualInterference,SecPandQPanalysis}, where we compare the derived bounds to their corresponding \gls{mc} simulations (Fig. \ref{FigTheoreticSetupSimulation}). The second is a practically oriented setup in which the \glspl{ms} are placed randomly in the plane with a more realistic channel model (cf. Fig. \ref{FigSystemSimulation}).

Beginning with the theoretical analysis setups,   Figures \ref{FigTheoreticSetupSimulation}(a) and \ref{FigTheoreticSetupSimulation}(b)
depict the standard-scheme performance (ergodic rate), evaluated via \gls{mc}
($10^{4}$ channel realizations), compared to the rate bound
 {described in \cref{remark_4}}.\footnote{ \label{FootnoteRateLB}{In cases where there is a closed-form expression for the rate under perfect \gls{csi}, the rate lower-bound follows by subtracting the rate gap from the perfect-\gls{csi} rate.}} 
Also included is the rate
under perfect \gls{csi} (cf. \eqref{eq_no_quanization_rate}). Considering
that transmitters could always turn off some of their antennas if it yields
a higher rate,  for each $B$, we picked  \({N}_{ t}\in\{2,\ldots,8\}\)
with the maximum rate.   The corresponding bound was also maximized over
$N_{t}$ for each $B$.

In Fig. \ref{FigTheoreticSetupSimulation}(a)  we considered   $Q=2$ \glspl{ms} 
placed  at   (-80,0) and (80,0) (in meters), served by  $M=4$ \glspl{srrh}. We  placed an \gls{srrh}
for each $ m\in\{1,\ldots
,M\}$ such that  its x-y coordinates   are the real and imaginary of $80
e ^{j\pi (1+2m)/4}$, respectively (in meters). Note that $\alpha_{q}= \sum_{m}\alpha_{q,m}$ is equal for $q\in\{1,2\}$. We used a  path-loss  exponent  of  3.5 and set the power
according to \cref{ass_power_standard_shceme} such that   $\alpha_{q} P_{\max}$
(cf. \cref{Theorem:Theorem1})  is 35 dB (black) and 15 dB (blue).  \cref{FigTheoreticSetupSimulation}(b)  considers a  symmetric network  that satisfies \cref{ass_equal_alphas} with $M=4$ \glspl{srrh} and $Q=8$ \glspl{ms} with a similar power allocation. The results show that the bound gets tighter as $B$ increases. Moreover, the bound exhibits the same behavior as the \gls{mc} simulation when $B$ increases.  {We note that the curves are not smooth since we allowed antenna turn off. The curve is unsmooth for bit numbers in which $N_{t}$ yielding the highest rate varies.}

Fig. \ref{FigTheoreticSetupSimulation}(c) compares the \gls{paq} and the standard schemes for the same setup as Fig.
\ref{FigTheoreticSetupSimulation}(b) with $P_{\max}=35$ dB; hence the black
curves are the same in both figures,  except of the horizontal logarithmic scale. 
We calculated the \gls{paq} rate bound using \cref{cor_P_and_Q_symmetric}, and evaluated the \gls{paq} scheme rate via  \gls{mc} ($10^{4}$ channel realizations)  where we    maximized it also over all feasible values of \(\bar Q.\) 
 The result indicates that the \gls{paq} scheme provides a significant performance gain; that is, $\hat{\tilde
        R }$ is much greater than $\hat R$
for at least 250 bits. Moreover, in the \gls{paq} scheme, the bound is tighter and approaches the \gls{mc} simulation way faster than the corresponding bound in the standard scheme. 

To further investigate the \gls{paq} scheme,
we study a practically oriented setup with randomly dispersed \glspl{ms} while considering propagation loss and shadowing. 
 The format includes a cluster of \(M = 4\) \glspl{srrh} creating a 100 m edge-length rhombus with an edge angle of 120$^{\circ}$. Each \gls{srrh}  is equipped with \(N_t = 8\)
isotropic transmit antennas. Eight  single-antenna \glspl{ms}  \((Q = 8)\)  were  placed uniformly at random
in the common area spanned by four hexagons, each one centered at a different \gls{srrh}. We set  a minimum distance of 10 m between each \gls{ms} and \gls{srrh}. The results were averaged over 20 realizations of \gls{ms}-placements, where each realization determined a set of attenuation factors \(\alphavec=\{\alpha_{q,m}:q=1\cdots8, m=1\cdots 4\}\) according     
\(\alpha_{q,m}=-128-37.6\log_{10}(r_{q,m})\) (in dB),\footnote{This model was used for urban-area non-line-of-site links  by the 3GPP; cf. page 61 3GPP
        Technical Report 36.814 \cite{attunuation2010}.} where  \(r_{q,m}\) is the distance
from S-RRH-\(m\) to MS-\(q\) in Km. The noise level at the receivers was  \(-121\) dBm.
For  each realization of \gls{ms}-placement, we calculated each \gls{ms} rate
by averaging over 40 channel realizations.  In  calculating  the network throughput, we averaged the rates of all \glspl{ms} across all placements. 
To ensure a fair comparison, we considered that the transmitters could always turn off some antennas to reduce the effective \gls{miso} channel dimensions.
Accordingly, in the standard scheme, we maximized the rate over $N_{\rm t}$, whereas, in the  \gls{paq} scheme, we maximized the rate over \(\bar Q\) while keeping $N_{t}= 8$. Finally, we set the overall power, transmitted to each \gls{ms}, fixed; i.e.,   $P_{q}=P_{q'}, \forall q,q'\in {\cal Q}$ (cf. \eqref{Signal-at-User-q}). To maintain  \(\Vert \pvec_{q} \Vert = 1\), each \gls{srrh} had to backoff its power 
until none was violating its individual power constraint \(P_{\rm max}\).\footnote{Note that while this power allocation strategy is not optimal, it yields good performance in high \glspl{snr}.} A more detailed description of this policy is given in (\cite{arad2020cran}).

 Fig. \ref{FigSystemSimulationA} presents
the throughput as a function of each \gls{srrh} transmit power, \(P_{\rm max}\) (cf. \cref{ass_power_constraint}).
 The results show that the \gls{paq} significantly outperformed the standard scheme. In the
latter, the network is already interference-limited  at 50 dBm, whereas in the former, at 110 dBm. Therefore, while the perfect-\gls{csi} throughput in the standard scheme is higher than the \gls{paq} counterpart, the latter goes up much faster. 

 Fig. \ref{FigSystemSimulationB} presents the average throughput as a function of $B$
under a per-\gls{srrh} power constraint
of   \(P_{\rm max}= 45\) dBm to study the effect of the quantization bits.  The result shows  that the
 \gls{paq} throughput  rapidly increases with 
$B$; thus, outperforming the standard scheme for a wide range of $B$. 

\section{Conclusions}
 \label{SecConclusions}

This article makes two contributions.
The first is a new upper bound on the
rate degradation experienced by  a cluster
of \glspl{srrh}, that employ  joint
 \gls{zf} with incomplete \gls{csi} compared to perfect \gls{csi}.
   The second is a new \gls{csi} sharing
scheme  that aims to reduce the \gls{csi}
overhead on the links between \glspl{srrh} in \gls{cran}.
The key distinguishing characteristic
of this scheme is that it applies
front-end matrices prior to \gls{csi}
quantization  to create
designated effective channels of low dimensionality; hence can
be quantized more accurately with fewer
bits. Furthermore, each \gls{srrh} serves
fewer \glspl{ms}, thus reducing
\gls{csi} and the number of  data streams delivered. We demonstrated,  through
analytical analysis and simulation,
that the proposed scheme achieves a
significant performance gain.

Possible extensions of this work would be  to optimize
the power allocation for each  \gls{ms} {and  optimize the dimension reduction level; i.e.,  $\bar Q$ (cf. \cref{Def_ms_allocation_stragegy}). Finally, it is necessary to explore channel models beyond independent Rayleigh fading.} 
\begin{appendices}
\section{} 
\label{App:proof of Lemma2:Bound
        on C}
To prove \cref{Lemma_Lemma2_BoundonC},
 we begin by rewriting the decomposition in \eqref{eq_DecompositionJindalhBar} as
\beq
\label{appendixA_eq1}
\bar\hvec_{q,m} = \hat{\bar\hvec}_{q,m}\cos\theta_{q,m} + \svec_{q,m} \sin\theta_{q,m},
\eeq
where $\bar\hvec_{q,m}=\hvec_{q,m}/\Vert \hvec_{q,m}\Vert$ and  $\hat{\bar\hvec}_{q,m}=\hat\hvec_{q,m}/\Vert \hat\hvec_{q,m}\Vert$, \(\theta_{q,m}\) is the angle between \(\bar\hvec_{q,m}\) and \(\hat{\bar\hvec}_{q,m}\), and \(\svec_{q,m}\in \comp^{ N_t\times 1}\) is a unit-norm random vector that is uniformly distributed over the null space of \(\hat{\bar\hvec}_{q,m}\) \cite{jindal2006mimo}.
Moreover, we define\footnote{Interchanging \(\bar\pvec_{j,m}\) and \(\hat{\bar\pvec}_{j,m}\) yields an equivalent decomposition of the quantized beamforming vector \cite{ravindran2008limited}; furthermore, \(\hat{\bar\pvec}_{j,m}\) is uniformly distributed.}
\beq
\label{appendixA_eq2}
\hat{\bar\pvec}_{j,m} = \bar\pvec_{j,m}\cos\phi_{j,m} + \gvec_{j,m} \sin\phi_{j,m}
\eeq
where $\bar{\pvec}_{j,m}=\pvec_{j,m}/\Vert \pvec_{j,m}\Vert$  and $\hat{\bar{\pvec}}_{j,m}=\hat\pvec_{j,m}/\Vert \hat\pvec_{j,m}\Vert$, \(\phi_{j,m}\) is the angle between \(\bar\pvec_{j,m}\) and \(\hat{\bar\pvec}_{j,m}\), and \(\gvec_{j,m}\in \comp^{ N_t\times 1}\) is a unit-norm random vector that is uniformly distributed over the null space of \(\bar\pvec_{j,m}\). Applying  Jensen's inequality to $A_{3}$, one obtains
\begin{IEEEeqnarray}{rcl}\label{eq_ Bound on C}
        A_3\leq \log \Big(1 +P\sum_{j\in\mathcal{Q}_{\mbox{\tiny-}q}}\E  \big\{  \vert{ \hvec^{\dagger}_q \hat
\pvec_j }\vert^2   \big\}\Big)
        \end{IEEEeqnarray}
and using \eqref{appendixA_eq1} and \eqref{appendixA_eq2}, the term with the expectation  in \eqref{eq_ Bound on C} can be written as
\begin{IEEEeqnarray}{rcl}\label{sum(|hat h p...|)}\Dcond{
                \E&&\big\{\vert\hvec^{\dagger}_q \hat\pvec_j \vert^2\big\}
                \\&&}{}
\begin{array}{ll}
                \Dcond{}{\E\big\{\vert\hvec^{\dagger}_q \hat\pvec_j\vert^2\big\}} =\E\Big\{\Big\vert\sum\limits_{m=1}^{ M}\bar\hvec^{ \dagger}_{q,m}\hat{\bar\pvec}_{j,m}\|\hvec _{q,m}\|\|\hat\pvec_{j,m}\|\Big|^2\Big\}=\E\Big\{\Big\vert\sum \limits_{m=1}^{ M}\Big(\hat{\hvec}^{ \dagger}_{q,m} \pvec_{j,m}{\Lambda}_{ 1,1_m}\\+\hat{\hvec}^{\dagger}_{q,m}\gvec_{ j,m}\|\hat\pvec_{j,m}\|{\Lambda}_{1,2_m}+\svec^{\dagger}_{q,m}\pvec_{j,m}\|\hvec _{q,m}\|{\Lambda}_{2,1_m}
                \Dcond{\\~~~~~~~~~~~~~~~}{}+\svec^{\dagger}_{q,m}\gvec_{j,m}\|\hvec _{q,m}\|\|\hat\pvec_{j,m}\|{\Lambda}_{ 2,2_m}\Big)\Big|^2\Big\}\end{array}
\end{IEEEeqnarray}
where ${\Lambda}_{k,l_{m}} =C_{k}(\theta_{q,m})C_{l}(\phi_{j,m}),$
$k,l \in\{1,2\}$ and  $C_{1}(\theta)=\cos \theta$,  $C_{2}(\theta)=\sin\theta$.
Extending \eqref{sum(|hat h p...|)}, one obtains
\begin{IEEEeqnarray}{rCl}\label{eq_ Exp(cdot)=D+E+F+G+H}
        \E\big\{\vert\hvec^{\dagger}_q \hat\pvec_j\vert^2\big\}=D + E
        + F + G + H
\end{IEEEeqnarray}
where
$
        D= \E \big\{\big\vert \sum_{m=1}^{M}\hat{\hvec}^{\dagger}_{q,m}
        \pvec_{j,m}{\Lambda}_{1,1_m}\big\vert^2\big\}$,  
        $
        E = \E  \big\{\big\vert{  \sum_{m=1}^{M}\hat{\hvec}^{\dagger}_{q,m}
                \gvec_{j,m}\|\hat\pvec_{j,m}\|{\Lambda}_{1,2_m} }\big\vert^2\big\}$,  
$ 
        F = \E  \big\{\big\vert  \sum_{m=1}^{M}$ ${\svec}^{\dagger}_{q,m} \pvec_{j,m}\|\hvec _{q,m}\|{\Lambda}_{2,1_m}\big\vert^2\big\}$
        $
        G = \E\big\{\big\vert  \sum_{m=1}^{M}{\svec}^{\dagger}_{q,m} \gvec_{j,m}\|\hvec _{q,m}\|\|\hat\pvec_{j,m}\|{\Lambda}_{2,2_m} \big\vert^2\big\}$
and   $  
 H =\sum_{i=1}^{12} \Xi_i$
in which  \(\{\Xi_i\}_{i=1}^{12}\)  includes all non-quadratic terms  
resulting from the extension of the  \gls{rhs} of \eqref{sum(|hat h p...|)}; i.e., terms that cannot be written as $\vert\cdot\vert^{2}$.
For example, $\Xi_1$ is given by
\begin{IEEEeqnarray}{rcl}
\label{eq_Xi=E(sum_sum))}
        \Xi_1
        = \E  \bigg\{\sum_{m=1}^{M}\sum_{n=1}^{M}\hat{\hvec}^{\dagger
        }_{q,m}\pvec_{j,m}\gvec^{\dagger
        }_{j,n}\hat{\hvec}_{q,n}\|\hat\pvec_{j,n}\|{\Lambda}_{1,1_m}{\Lambda}_{1,2_n}\bigg\}~~~~~
\end{IEEEeqnarray}
\begin{proposition}
  \label{PropBoundOnD}
The term $D$, in \eqref{eq_ Exp(cdot)=D+E+F+G+H}, satisfies
        \begin{IEEEeqnarray}{rcl}
        \label{eq_bound_D}
                D\leq \frac{{  \alpha_{q}}N_t}{M}\big((1-{\cal U}(2^{B/Q},a))^{2}-(1-{\cal U}(2^{B/Q},a)/2-{\cal
U}(2^{B/Q},2a))^{4}\big) 
\end{IEEEeqnarray}
where \(a=\frac{1}{N_{t}-1}\) and \({\cal
U}(\cdot)  \) is defined in \eqref{eq_Define_U}.
\end{proposition}
\begin{IEEEproof}
        Because $\{\Lambda _{1,1_m}\}_{m\in {\cal M}}$  are identically distributed, it can be written as $                \Lambda_{1,1_{m}} = \sigma +{\bar\Lambda}_{1,1_{m}},  \forall m\in{\cal M}$, 
        where \(\E \{{\bar\Lambda}_{1,1_{m}}\} = 0\) and \(\sigma=\E \{{\Lambda}_{1,1_{m}}\} \). Substituting 
      $\Lambda_{1,1_{m}}$ into $D$ (cf. \eqref{eq_ Exp(cdot)=D+E+F+G+H}) while recalling  that  $\{\theta_{q,i},\phi_{j,i}\} _{                i\in {\cal M}, q,j\in{\cal
                        Q}}$, are independent of 
        \(\{\hat{\hvec}_{q,i},\pvec_{j,i}\}_{i\in {\cal M},
                q,j \in{\cal Q}}\), it can be shown that
        \begin{IEEEeqnarray}{rcl}
                D
                &=&\sigma^2\E\Big\{\big|\hat{\hvec}^{\dagger}_{q}\pvec_{j}\big|^2\Big\}+\E\Big\{\Big|\sum_{m=1}^{M} \hat{\hvec}^{\dagger}_{q,m}\pvec_{j,m}{\bar\Lambda}_{1,1_{m}}\Big|^2\Big\} \label{eq_DoubleSum D}\\
                && + \sigma\sum_{m=1}^{M}\sum_{n=1}^{M}\E \bigg\{\hat{\hvec}^{\dagger}_{q,m}\pvec_{j,m}\pvec_{j,n}^{\dagger}\hat{\hvec}_{q,n}\bigg\}\E
                        \left\{{\bar\Lambda}_{1,1_{n}} \right\}\Dcond{\nonumber\\&&}{}+ \sigma\sum_{m=1}^{M}\sum_{n=1}^{M}\E \bigg\{\hat{\hvec}^{\dagger}_{q,m}\pvec_{j,m}\pvec_{j,n}^{\dagger}\hat{\hvec}_{q,n}\bigg\}\E \left\{{\bar\Lambda}_{1,1_{m}} \right\}
                \nonumber\\
                &=&
                 \E \bigg\{\Big|\sum_{m=1}^
                {M}\hat{\hvec}^{\dagger}_{q,m}\pvec_{j,m}{\bar\Lambda}_{1,1_{m}}\Big| ^2\bigg\}\label{eq_expression_for_D}
        \end{IEEEeqnarray}
        To obtain the latter, we    also  used  $\hat\hvec_q^{\dagger}\pvec_j=0,\;
        \forall j\neq q\in{\cal Q}$ and \(\E \{{\bar\Lambda}_{1,1_{m}}\} = 0\).
Hence
        \begin{IEEEeqnarray}{rcl}\label{eq_D_equal_I+JLine1}
                D&=&\E  \bigg\{\sum_{m=1}^{M}\big\vert{\hat{ \hvec}^{\dagger}_{q,m} \pvec_{j,m}}\big\vert^2{\bar\Lambda}^2_{1,1_{m}}\bigg\}\Dcond{\\\nonumber
                        &&\;\;}{}+\E  \bigg\{\sum_{m=1}^{M}\sum_{n\neq m}^{M}\hat{\hvec} ^{\dagger}_{q,m}\pvec_{j,m}\pvec_{j,n}^{\dagger}\hat{\hvec}_{q,n}{\bar\Lambda }_{1,1_{m}}{\bar\Lambda}_{1,1_{n}}\bigg\}\\
                &=&\sum_{m=1}^{M}\underbrace{\E \big\{\|\pvec_{j,m}\|^2\|\hat{\hvec}_{q,m}\|^2\big\vert{ \hat{\bar\hvec}^{\dagger}_{q,m}\bar\pvec_{j,m}}\big\vert^2\big\}}_{I_{q,m}
                } \underbrace{\E\left\{{\bar\Lambda}^2_{1,1_{m}}\right\}}_{J}\label{eq_D_equal_I+J}
        \end{IEEEeqnarray}
        where we again used  the independence between $\{\theta_{q,i},\phi_{j,i}\}_{i\in {\cal M}, q,j \in{\cal Q}}$ and \(\{\hat{\hvec}_{q,i},\pvec_{j,i}\}_{i\in {\cal M},
                q,j \in{\cal Q}}\) as well as the independence between ${\bar\Lambda }_{1,1_{m}}$ and ${\bar\Lambda}_{1,1_{n}} \forall m\neq n\in{\cal M}$.
        
        Next, using  $I_{q,m}=\sqrt{I^2_{q,m}}$ and applying
        the Cauchy-Schwarz inequality, 
        one obtains
        \begin{IEEEeqnarray}{rcl}
        \label{eq_bound_I_CS}
                I_{q,m}&&\leq\sqrt{\E^2\big\{\|\pvec_{j,m}\|^2\big\}\E^2 \big\{\|\hat{\hvec}_{q,m}\|^2\big\}\E^2 \big\{\vert{\hat{\bar\hvec}^{\dagger}_{q,m}\bar\pvec_{j,m}}\vert^2\big\}}\Dcond{\nonumber\\
                        &&}{}\overset{({\rm a})}=\frac{\alpha_{q,m}N_t}{M}\E \big\{\big\vert{\hat{\bar\hvec}^{\dagger}_{q,m}\bar\pvec_{j,m}}\big\vert{}^2\big\}\overset{({\rm b})}\leq \frac{\alpha_{q,m}N_t}{M}\;\;\;
        \end{IEEEeqnarray}
        wherein (a) we used \cref{ass_channelModel},  \(\E\{\|\hat{\hvec}_{q,m}\|^2\}/ \alpha
                _{q,m}
        = N_t\) \cite{jindal2006mimo}, and \(\E \{\|\pvec_{j,m}\|^2\} = \frac{1}{M}\). The latter follows  because    \(\|\pvec_{j}\|^2
        = 1 \) and  \(\{\|\pvec_{j,m}\|\}_{m\in {\cal M}}\) are identically distributed. In (b) we used \(\E\{\big\vert{\hat{\bar\hvec}^{\dagger}_{q,m}\bar\pvec_{j,m}}\big\vert{}^2\big\}\leq 1\).
        Proceeding to $J$ (cf. \eqref{eq_D_equal_I+J}), note that
\begin{IEEEeqnarray}{rcl}
\label{appendixA_eq11}
J&=&\E\{(\Lambda_{1,1_{m}}-\E \{\Lambda_{1,1_{m}}\} )^2\}
                \Dcond{\nonumber\\&=&}{=} \E\{\cos^2\theta_{q,m}\cos^2\phi_{j,m}\}-(\E  \{\cos\theta_{q,m}\cos\phi_{j,m}\}) ^2\nonumber\\
                &=& \E^{2}\{\cos^2\theta_{q,m}\}-\E^{4}  \{\cos\theta_{q,m}\},  \end{IEEEeqnarray}
        where the latter  follows since \(\theta_{q,m}\) and \(\phi_{j,m}, \ \forall m\in{\cal M  },\) \(q,j\in{\cal Q  }\), are independent   identically distributed.
        Because \( \hvec_{q,m}\) is quantized with \(B/Q\) bits, it follows that 
        \cite{jindal2006mimo}
        \begin{IEEEeqnarray}{rcl}\label{eq_Exp_cos_2_standard}
                \E\left\{\cos^2\theta_{q,m}\right\}=1-{\cal U}\big(2^{B/Q},a\big)
        \end{IEEEeqnarray}
        where ${\cal U}$ is defined in \eqref{eq_Define_U}. Before continuing,              
note  that
\begin{IEEEeqnarray}{rCl}
\label{eq_inequality_cos_theta}
\E\left\{\cos\theta_{q,m}\right\}\geq 1-{\cal U}(2^{B/Q},a)/2-{\cal U}(2^{B/Q},2a)
\end{IEEEeqnarray}
 where we used $\cos\theta_{q,m}= \sqrt{1-\sin{}^{2}(\theta_{q,m})}$
and  the inequality $\sqrt{1-x}\geq 1-x/2-x^{2},\; \forall x\in[0,1]$. Thus, 
$J\leq (1-{\cal U}(2^{B/Q},a))^{2}-(1-{\cal U}(2^{B/Q},a)/2-{\cal
U}(2^{B/Q},2a))^{4}
,$
 \space which together with \(\sum_{m=1}^M \alpha_{q,m}=\alpha_{q}\) (cf.  \cref{Theorem:Theorem1}) establishes the desired result.
\end{IEEEproof}
\begin{proposition} \label{Prop_Bound_on_E_F}
        The terms $E$ and $F$, given in \eqref{eq_ Exp(cdot)=D+E+F+G+H},  satisfy  \begin{IEEEeqnarray}{rcl}
\label{eq_inqaulity_E_F_standard}
E,F\leq \frac{ {  \alpha_{q}}N_t}{M(N_t - 1)}\left(1-\mathcal{U}\left(2^{B/Q},a\right)\right)2^{\frac{-B}{Q(N_t - 1)}} \end{IEEEeqnarray}
\end{proposition}
\begin{IEEEproof}
        Rewriting $E$ (cf. \eqref{eq_ Exp(cdot)=D+E+F+G+H}) one obtains
        \begin{IEEEeqnarray}{rcl}\label{eq_E_equals_standard}
          E&=&\E\bigg\{\sum_{m=1}^{M }\big\vert{\hat{\hvec}^{\dagger}_{q,m}\gvec_{j,m}\Vert \hat \pvec_{j,m} \Vert}\big\vert^2{\Lambda}^2_{1, 2_m}\bigg\}\Dcond{\\\nonumber &&}{\\&&}+\E \bigg\{\sum_{m=1}^{M }\sum_{n\neq m}^{M} \hat{\hvec}^{\dagger}_{q,m}\gvec_{j,m}\gvec^{ \dagger}_{ j,n}\hat{\hvec}_{q,n}\Vert \hat \pvec_{j,m} \Vert\Dcond{\\&&~~~~~~~~~~~~~~~~~\times}{}\Vert \hat \pvec_{j,n} \Vert{\Lambda}_{1,2_m}{\Lambda}_{1,2_n} \bigg\} \label{appendixA_eq16SecondLine}
        \end{IEEEeqnarray}
        Now, denote
        \begin{IEEEeqnarray}{rcl}\label{defenition_of_u}
                w = \hat{\bar\hvec}_{q,m},  \gvec_{j,n},\hat{\bar\hvec}_{q,n},\|\hvec_{q,m}\|,\|\hvec_{q,n}\|,\|\hat \pvec_{j,m}\|,\|\hat \pvec_{j,n}\|
        \end{IEEEeqnarray}
        and using the same independence argument  as in \eqref{eq_DoubleSum D}, the double sum in \eqref{appendixA_eq16SecondLine} can be written as
        \begin{IEEEeqnarray}{rcl}
                &&\sum_{m=1}^{M }\sum_{n\neq m}^{M} \E \Big\{\E \Big\{\hat{\hvec}^{\dagger}_{q,m}\gvec_{j,m}\gvec^{ \dagger}_{ j,n}\hat{\hvec}_{q,n}\Vert \hat \pvec_{j,m} \Vert\Vert \hat \pvec_{j,n} \Vert\Big| \bar\pvec_{j,m},w \Big\}\Big\}\Dcond{\nonumber\\
                        &&~~~~~~~~~~~~\times}{}\E \Big\{{\Lambda}_{1,2_m}{\Lambda}_{1,2_n} \Big\}\label{double_sum_with_total_expectation}
        \end{IEEEeqnarray}
        Given $w$, all the arguments inside the internal expectation are constants, except $\gvec_{j,m}$. Furthermore,
        recalling  that given $\bar\pvec_{j,m}$, $\gvec_{j,m}$ is uniformly distributed
        on the unit sphere of the null space of  $\bar\pvec_{j,m}$, it follows that $\E\big\{\gvec_{j,m}\big| \bar\pvec_{j,m},w\big\}=\0_{ N_t}.$
          Thus, the double sum in \eqref{appendixA_eq16SecondLine}
        is equal to zero.  Applying the  Cauchy-Schwarz inequality to  \eqref{eq_E_equals_standard} and using the independence argument  again  as in \eqref{eq_DoubleSum D}, one obtains \begin{IEEEeqnarray}{rcl}
                \label{AppendixA_eq16ThirdLilne} E &\leq&
                \sum_{m=1}^{M} \E  \big\{\big\vert{\hat{\bar\hvec}^{\dagger}_{q,m} \gvec_{j,m}}\big\vert^2\big\}\E
                \big\{\|\hvec_{q,m}\|^{2}\big\}\Dcond{\nonumber \\}{}\label{AppendixA_eq16FourthLilne} \Dcond{&&~~~~~~~~\times}{}\E\big \{\|\hat\pvec_{j,m}\|^2\big\}\E\big\{{ \Lambda}^2_{1,2_m} \big\}\;. \end{IEEEeqnarray}
        Next, from 
 \eqref{AppendixA_eq16FourthLilne},
\begin{IEEEeqnarray}{rcl}
\label{eq_E_inequality_second_standard}
                E&\leq&\frac{{  \alpha_{q}}N_t}{M} \E\Big\{\big\vert{\hat{\bar\hvec}^{\dagger}_{q ,m}\gvec_{j, m}}\big\vert^2\Big \}\Dcond{\\&&\times}{}\E\Big\{\cos^2\theta_{q,m}\Big\}\E\Big\{\sin^2\phi _{j,m}\Big\}\label{appendixA_eq17_2}
\end{IEEEeqnarray}
        where we used similar arguments as in \eqref{appendixA_eq11} concerning the angles, and in addition,  \(
\E\{\|\bar\hvec_{q,m}\|^2\}/ \alpha
                _{q,m}= N_t\) \cite{jindal2006mimo}, \(\sum_{m=1}^M \alpha_{q,m}=\alpha_{q}\), and \(\E \left\{\|\hat\pvec_{j,m}\|^2 \right\} = \frac{1}{M}\).  

To further simplify \eqref{eq_E_inequality_second_standard},  we treat each of  the expressions in the \gls{rhs}  separately. First
        \begin{IEEEeqnarray}{rcl}
                \label{Ecos^2(theta)ESin^2(phi)}
                \E  \left\{\cos^2\theta_{q,m}\right\}\E\left\{\sin^2\phi_{j,m}\right\} \leq \left(1-\mathcal{U}\left(2^{B/Q},a\right)\right)2^{\frac{-B}{Q(N_t - 1)}}
        \end{IEEEeqnarray}
        where   we used  \eqref{eq_Exp_cos_2_standard}
        as well as the upper bound  \cite{jindal2006mimo} \beq\E\{\sin^2\phi_{j,m}\}\leq 2^{\frac{-B}{Q(N_t -1)}}.\label{Eq_E[sin2(phi_(j,m))]}\eeq
Next, consider $            \hat{\bar\hvec}_{q,m} = \Pmat_{\gvec_{j,m}}\hat{\bar\hvec}_{q,m} + \Pmat^{\perp}_{\gvec_{j,m}}\hat{\bar\hvec}_{q,m} $,
        where \(\Pmat_{\gvec_{j,m}}\), \(\Pmat^{\perp}_{\gvec_{j,m}}\)
 are the projection matrices into  space spanned by  \(\gvec_{j,m}\) and  its orthogonal complement, respectively.  It follows that
        \begin{IEEEeqnarray}{rcl}\label{appendixA_eq20}
                \E&&\Big\{\vert{\hat{\bar\hvec}^{\dagger}_{q,m}\gvec_{j,m}}\vert^2\Big\}\Dcond{\nonumber\\ &&}{}=\E\big\{\big\vert{\big((\Pmat_{\gvec_{j,m}}\hat{\bar\hvec}_{q,m})^{\dagger} + (\Pmat^{\perp}_{\gvec_{j,m}}\hat{\bar\hvec}_{q,m})^{\dagger}\big)\gvec_{j,m}}\big\vert^2\big\}\nonumber\\
                &&\Dcond{}{~~~~~} \overset{({\rm a})}= \E\big\{\big\vert{({\Pmat_{\gvec_{j,m}}\hat{\bar\hvec}_{q,m}})^{\dagger} \gvec_{j,m}\big\vert^2}\big\}\Dcond{\nonumber\\
                        &&}{}\overset{({\rm b})}\leq \E\bigg\{\bigg\vert
                \mbox{$
                        \frac{({{\Pmat_{\gvec_{j,m}}\hat{\bar\hvec}_{q,m}})^{\dagger}} }{\Vert \Pmat_{\gvec_{j,m}}\hat{\bar\hvec}_{q,m}\Vert}
                        $}
                \gvec_{j,m}\bigg \vert^2\bigg\} \overset{({\rm c})}= \frac{1}{N_t - 1}
        \end{IEEEeqnarray}
        where (a) follows because $\Pmat_{\gvec_{j,m}}^{\perp}\gvec_{j,m}=\0_{ N_t}$ and  (b) follows  from \(\|\Pmat_{\gvec_{j,m}}\hat{\bar\hvec}_{q,m}\| \leq 1\) (recall that  \(\|\hat{\bar\hvec}_{q,m}\|
        = 1\)).  (c)  follows because  \(\gvec_{j,m}\) is independent of \({\Pmat_{\gvec_{j,m}}\hat{\bar\hvec}_{q,m}}\), and is uniformly distributed on the unit sphere of the \((N_t - 1)\)-dimensional null space of \(\bar{\pvec}_{j,m}\). Thus, the expectation on the  left-hand side of (c) is taken according to the  \(\beta(1,N_t - 2)\) distribution \cite{jindal2006mimo}.
        
 Substituting   \eqref{Ecos^2(theta)ESin^2(phi)} and \eqref{appendixA_eq20} into \eqref{eq_E_inequality_second_standard}  establishes   \eqref{eq_inqaulity_E_F_standard}  for $E$. The proof for $F$ is identical and is omitted here due to space limitations.
\end{IEEEproof}
\begin{proposition}
\label{PropBoundOnG}
        The term $G$ in \eqref{eq_ Exp(cdot)=D+E+F+G+H} satisfies $  G\leq\   \frac{ {  \alpha_{q}}N_t}{M(N_t - 1)}2^{-\frac{2B}{Q(N_t - 1)}} $.
\end{proposition}
\begin{IEEEproof} Similar to the derivation of \eqref{AppendixA_eq16ThirdLilne}, it can be shown that 
        \begin{IEEEeqnarray}{rcl}
                G& \leq& \sum_{m=1}^{M} \E  \Big\{\big\vert{{\svec}^{\dagger}_{q,m}\gvec_{j,m}}\big\vert^2\Big\}\E\big\{ \|\hvec_{q,m}\|^{2}\big\}\Dcond{\nonumber\\&&~~~~~~~\times}{}\E\big\{\|\hat\pvec_{j,m}\|^2\big\}\E \left\{{\Lambda}^2_{2,2_m} \right\}\;.\label{appendixA_eq22}
        \end{IEEEeqnarray}
        Next,  \(\E\{\vert{\svec^{\dagger}_{q,m}\gvec_{j,m}}\vert^2\}\) can be bounded using similar arguments as in
        \eqref{appendixA_eq20}, and by further employing  \eqref{Eq_E[sin2(phi_(j,m))]}, one obtains the desired result. 
\end{IEEEproof}
\begin{proposition}\label{PropBoundOnH} Consider 
        $H=\{\Xi_{i}\}_{i=1}^{12}$ (cf.  \eqref{eq_ Exp(cdot)=D+E+F+G+H}), then $              \Xi_i = 0,\; \forall i\in\{1 \ldots 12\}$.
\end{proposition} \begin{IEEEproof}
        The proposition will be proven only for
        $\Xi_{1}$, where for the rest $\Xi_{i}$, $i>1$ the proof is identical.
%
Similar to \eqref{double_sum_with_total_expectation}, and with $w = \hat{\bar\hvec}_{q,m}, \bar \pvec_{j,m},\hat{\bar\hvec}_{q,n},\|\hvec_{q,m}\|,\|\hvec_{q,n}\|,\|\hat \pvec_{j,m}\|,\|\hat \pvec_{j,n}\|$
        it can be shown that $\E\big\{\gvec^{\dagger
        }_{j,n}\big| \bar\pvec_{j,n},w\big\}=\0_{ N_t}.$ 
        Thus, $\Xi_1 =0$, which establishes the desired result. 
\end{IEEEproof}
To complete the proof, we apply  \cref{PropBoundOnD,Prop_Bound_on_E_F,PropBoundOnG,PropBoundOnH} on \eqref{eq_ Exp(cdot)=D+E+F+G+H}, and in turn, substitute the result in \eqref{eq_ Bound on C}, which establishes the desired result.

\section{}
\label{AppendixProofLemma3}
To prove  Lemma \ref{Lemma:Lemma3 Bound A-B}, we begin by bounding  $A_2$ (cf. \eqref{eq_Delta_R_with_A1A2A3theorem_1}). Let $\theta_{q}=\angle \langle \bar\hvec_{q},
\hat{\bar{\hvec}}_q\rangle$ where
$\bar\hvec_{q}=\hvec_{q}/\Vert\hvec_{q}\Vert$
(cf. \eqref{eq_define hq}), $\hat{\bar\hvec}_{q}=\hat\hvec_{q}/\Vert\hat\hvec_{q}\Vert$
(cf. \eqref{eq_define hq hat}) and    $\xi_{q}=\angle \langle
\hat{\bar{\hvec}}_q,\hat{\pvec}_q\rangle$. Since the argument of the logarithm in $A_{2}$ is always
positive; i.e.,
\(\vert \bar\hvec^{\dagger}_q \hat\pvec_q \vert >0\), we assume
without loss of generality that \(\angle\langle\bar\hvec_q,\hat\pvec_q\rangle \in [0,\pi/2]\).
Employing the triangle inequality
for angles, one obtains,  \begin{IEEEeqnarray}{rCl}
        \angle \langle\bar\hvec_q,\hat\pvec_q\rangle \leq \theta_q +\xi_q
        \label{eq_tr_ineq_angles}
\end{IEEEeqnarray}
Applying  $\cos^{2}(\cdot)$
on both sides is possible if  
$ 0\leq
\xi_{q}+\theta
_q\leq \frac{\pi}{2}$, which  guarantees the monotonicity of the cosine. Denote the common
probability space, in which all random variables we are dealing with are defined, by
$(\Omega,{\cal F},{\mathbb P}),$ and let
\begin{IEEEeqnarray}{rCl}
        \label{eq_define_Acal_q}
        \mathcal{A}_q=\big\{\omega\in\Omega:0\leq
        \xi_{q}(\omega)+\theta _q(\omega)\leq \frac{\pi }{2}\big\}.
\end{IEEEeqnarray}
Furthermore, for $X$,  a random variable defined on $(\Omega,{\cal
        F},{\mathbb P})$ and  for ${\cal A}\in {\cal F}$, we denote the random
variable $\chi_{{\cal A}}(\omega)X(\omega)$ by $ \chi_{{\cal
                A}}X$. Combining  \eqref{eq_define_Acal_q}  and  \eqref{eq_tr_ineq_angles}, it follows that 
\begin{IEEEeqnarray}{rcl}\label{jcos2(jalgle jbar h q,jhat p
                q)jleqfirst1}
        \vert \bar\hvec^{\dagger}_q
        \hat\pvec_q \vert^2 =\cos^2\big(\angle\langle\bar\hvec_q,\hat \pvec_q\rangle\big)
        \Dcond{&\geq&}{&\geq& }\chi_{{\cal A}_{q}}\cos^2\big(\angle\langle\bar\hvec_q, \hat\pvec_q\rangle\big)
        \geq \chi_{{\cal A}_{q}}\cos ^2\big(\xi_{q}
        + \theta_q\big)\end{IEEEeqnarray}
and from the definition of $A_{2}$ (cf. \eqref{eq_Delta_R_with_A1A2A3theorem_1}) and the monotonicity of \(\log(1+x)\),  it follows that 
\begin{IEEEeqnarray}{rCl}
        \label{ep_bound_on_A_2}         A_2&\geq& 
        \E\big\{\chi_{{\cal A}_{q}}\log \big(1+P \|\hvec_q\|^2\cos^{2}(\xi_{q}+\theta_{q})
        \big)\big\}.
\end{IEEEeqnarray}
\begin{proposition} 
Let
        $
        K(x)\define\frac{\pi
                x  }{2 \sqrt{x+1}}$, then
        \begin{IEEEeqnarray}{rCl}
        \label{eq_LowerBound_Lipschitz}
                \log \big(1+P \|\hvec_q\|^2\cos^{2}(\xi_{q}+\theta_{q})\big)\geq
                \log \big(1+P \|\hvec_q\|^2 \cos ^2(\xi _{q})\big)-\sin(\theta_{q}) K(P \|\hvec_q\|^2).
        \end{IEEEeqnarray}  
\end{proposition}
\begin{IEEEproof}
        {Let $\check P=P\Vert\hvec_{q}\Vert^{2}$ and 
                $g(\theta ,\xi )\define \log (1+ \check P \cos^2(\theta
                +\xi ))$
                for $\theta,\xi\in[0,\pi/2],$ where for brevity we omit the subscript $q$ in this proof. To bound $g(\theta,\xi)$, we solve $g^{(2,0)}(\theta ,\xi )=0,$  from which the inflection point is given by
                $ \theta _{\text{I}}=\cos ^{-1}({({1}/{\check P+2})^{1/2}})-\xi$
                and $ g^{(1,0)}(\theta_{\rm I} ,\xi )=-{\check P}{{(\check P+1)^{-1/2}}}$. Furthermore, it can be shown
                that  $g(\theta ,\xi )$  is convex for   $\theta >\theta _{\text{I}}$
                and concave otherwise.
                We first derive a bound for the case where
                $\theta_{\rm I}\leq 0$, in which   
                $g(\theta ,\xi )$  is convex for $\theta\in[0,\pi/2]$ and
                therefore, 
                $g(\theta ,\xi )\geq \theta  g^{(1,0)}(0,\xi )+g(0,\xi )
                $.  Moreover, it can be shown that   $g^{(1,0)}(\theta ,\xi )<0, \forall\theta
                +\xi <\frac{\pi }{2}$. Then, using  $\frac{\pi}{2}  \sin (\theta )>\theta$
                we replace  $\theta$ with $\frac{\pi}{2}\sin(\theta)$ and obtain
                $g(\theta ,\xi )\geq \frac{\pi}{2}   \sin (\theta ) g^{(1,0)}(0,\xi
                )+g(0,\xi)$. Noting that  $
                g^{(1,0)}(0,\xi )=-\frac{2 \check P \sin (\xi ) \cos (\xi )}{\check P \cos
                        ^2(\xi )+1}\geq -{\check P}{{(\check P+1)^{-1/2}}}
                $ and  substituting $g(0,\xi )$ into the latter inequality, we obtain \eqref{eq_LowerBound_Lipschitz}, which establishes the result for $\theta _{\text{I}}\leq0$.
                In  the case where $\theta _{\text{I}}>0$, we use the Lipschitz continuity; i.e.,  a function $f(x)$ is Lipschitz continuous if $\exists C>0$ such that
                $\left| f\left(x _1\right)-f\left(x _2\right)\right|\leq C \left| x _1-x _2\right|
                $ $\forall x_{1},x_{2}$. If $f(x)$ is differentiable, then   $C=\sup _{x}\left\vert {d f(x)}/{dx}\right\vert$. Since $g(\theta ,\xi )$ is Lipschitz continuous, by
                replacing
                \(C\) with  the inflection point, one
                obtains
                $ 
                g(\theta ,\xi )\geq g(0,\xi )+\frac{\pi}{2}   \sin (\theta
                ) g^{(1,0)}\big(\theta _{\text{I}},\xi \big)=\log \big(1+ \check P \cos
                ^2(\xi )\big)-\sin(\theta) K(\check P).$ Finally, because
                the bounds  for $\theta _{\text{I}}\leq0$ and $\theta _{\text{I}}>0$ are identical, the desired result follows.}
\end{IEEEproof}

Substituting \eqref{eq_LowerBound_Lipschitz}  into   \eqref{ep_bound_on_A_2}, it follows that \beq  \label{eq_bound_A_2_useing_Appendix_inequality}
        A_2\geq E_{2}-       \E\big\{ \chi_{{\cal A}_{q}}\sin(\theta_q) K(P
\|\hvec_q\|^2)
        \big\}\eeq
 where $E_{2}=\E\big\{
\chi_{{\cal A}_{q}}\log\big(1+P
\|\hvec_q\|^2 \cos^2\xi _{q}
\big) \big\}$, and  by  Cauchy–Schwarz and    Jensen's inequality ($K$ is concave) it follows
\(
  A_2\geq E_{2}-   \E\{\chi_{{\cal A}_{q}}\}\E\{\sin(\theta_q)\}
K(P
\E\{\|\hvec_q\|^2
        \}).\)
Hence 
\begin{IEEEeqnarray}{rcl}
        \label{eqA2BoundJenssionAppB}
        A_2&\geq &E_{2}-\E\{\chi_{{\cal A}_{q}}\}\E\{\sqrt{Z_{q}}\}\frac{\pi
                \alpha_{q} P N_{t} }{2 \sqrt{\alpha_{q}PN_{t}+1}}
\end{IEEEeqnarray}
where  we substituted $\sin(\theta_{q})=\sqrt {Z_{q}}$. 
To bound the term  $E_{2}$ (cf. \eqref{eq_bound_A_2_useing_Appendix_inequality}) we   invoke, once again, the triangular inequity for angles, $\xi _{q}
\leq \phi_q + \angle \langle\hat{ \bar\hvec}_q,\pvec_q\rangle,$
where  $\phi_q=\angle\langle \pvec_q,\hat{\pvec}_q \rangle$, and obtain
\begin{IEEEeqnarray}{rCl}
        \label{ep_bound_on_E_2} 
        E_2&\geq& 
        \E\Big\{\chi_{{\cal C}_{q}}
        \log \Big(1+P \Vert\hvec_{q}\Vert^{2}\cos ^2\big(\angle  \big\langle \hat{\bar{{\hvec}}}_q,
        {\pvec}_q\big\rangle +\phi _q
        \big)\Big)\Big\}
\end{IEEEeqnarray}
where  $\mathcal{C}_q={\cal A}_{q}\cap{\cal B}_{q} $ and  $ {\cal B}_q=\big\{\omega\in\Omega:0\leq\angle
\big\langle \hat{\bar{{\hvec}}}_q,
{\pvec}_q\big\rangle(\omega) +\phi_q(\omega) \leq\frac\pi 2\big \}$.
Using similar arguments as in  \eqref{eq_LowerBound_Lipschitz}-\eqref{eqA2BoundJenssionAppB}, it can be shown that
\begin{IEEEeqnarray}{rCl}
        E_2&\geq&\E\big\{\chi_{{\cal C}_{q}}\log \big(1+P \Vert \hvec_{q} \Vert^{2} \cos
        ^2(\angle \langle \hat{\bar{{\hvec}}}_q,{\pvec}_q
        \rangle )\big)\big\}
-\E\{\chi_{{\cal C}_{q}}\}\E\{\sin(\phi_{q})\}\frac{\pi
                \alpha_{q} P N_{t} }{2 \sqrt{\alpha_{q}PN_{t}+1}}
        \label{eq_B1<=}\end{IEEEeqnarray}

Before continuing, the following proposition is necessary.
\begin{proposition}
        \label{Propbound on theta_q}
        Let   $Z_{q}=\sin^{2}(\theta_{q})$.
        Then, \(Z_{q}^{\min}\leq Z_{q}\leq
        Z_{q}^{\max}\), where $Z_{q}^{\min}=\min\{Z_{q,l},l\in{\cal M}\}$
        and $Z_{q}^{\max}=\max\{Z_{q,l},l\in{\cal M}\}$. Moreover, 
$                {\cal U}
                (2^{B/Q},a/2)-{\cal V}_{M}(2^{B/Q},a/2)\leq\E\{\sqrt{Z_{q}}\}\leq
                {\cal U}(2^{B/Q},a/2)+{\cal V}_{M}(2^{B/Q},a/2)
$
        \space where \({\cal
                U}(\cdot)  \) and \({\cal V}_{M}(\cdot)  \) defined in \eqref{eq_Define_U} and \eqref{eq_define_U2}, respectively.
        \label{Prop_Sin_th_q}
\end{proposition}
\begin{IEEEproof} 
        Consider 
$
                Z_{q}=1- \frac{\vert\hat{\hvec}_q^{\dagger}
                        \hvec_q\vert{}^2}{\Vert
                        \hvec_q  \Vert^{2}\Vert \hat{\hvec}_q \Vert^{2}}
                =\sum _{u=1}^M \sum _{l=1}^M \frac{\Vert
                        \hvec_{q,l} \Vert{}^2 \Vert
                        \hat{\hvec}_{q,u} \Vert{}^2}{\Vert
                        \hvec_{q} \Vert{}^2 \Vert\hat \hvec_{q} \Vert {}^2}
                \Big(1- \sqrt{1-Z_{q,l}}
                \sqrt{1-Z_{q,u}}\Big)$
        where  we used \eqref{eq_define hq}, \eqref{eq_define hq hat}
        and \eqref{eq_DecompositionJindalhBar}. Thus,
        \begin{IEEEeqnarray}{rCl}
                \nonumber        Z_{q}&\leq&\frac{1}{\Vert \hvec_q \Vert{}^2
                        \Vert
                        \hat{\hvec}_q \Vert{}^2}\sum _{u=1}^M \sum _{l=1}^M
                \Vert
                \hvec_{q,l} \Vert{}^2 \Vert \hat\hvec_{q,u} \Vert {}^2\Dcond{
                        \nonumber\\&& \times}{}
                \Big(1-\big(\min\big\{ \sqrt{1-Z_{q,l}},l\in{\cal M}\big
                \}\big)^{2}
                \Big)       \label{Bound on Sin theta}
                =
                Z_{q}^{\max}
        \end{IEEEeqnarray}
        Similarly, it can be shown that  $Z_{q}\geq Z_{q}^{ \min},$
        which establishes the first statement of the theorem. The bound on $\E\{\sqrt{Z_{q}}\}$  follows from the
        identities, 
        \(
        \E\{Z_{q}^{\max}\}\leq  \E\{Z_{q,l}\}  +\frac{(M-1) \sqrt{
                        {\rm var}\{Z_{q,l}\}}}{  \sqrt{2 M-1}}
        \),  \(
        \E\{Z_{q}^{\min}\}\geq \E\{Z_{q,l}\}-\frac{(M-1) \sqrt{
                        {\rm
                                var}\{Z_{q,l}\} }}{  \sqrt{2 M-1}}
        \) (cf. \cite{David2003a},
        Sec. 4.2).  
\end{IEEEproof}

Note that $\sin(\theta_{q})$ and $\sin(\phi_{q})$ are identically distributed; therefore, the expression in Proposition \ref{Prop_Sin_th_q} also applies to $\sin(\phi_{q})$.

Returning  to the main proof; 
by invoking Proposition \ref{Prop_Sin_th_q} 
and combining  the latter with  \eqref{eqA2BoundJenssionAppB} and  \eqref{eq_B1<=}, one obtains
$        A_2\geq
         \E\{\chi_{{\cal C}_{q}} \log( 1+P\vert\hat\hvec^{\dagger}_q\pvec_q
        \vert^2)\}-(\E\{\chi_{{\cal A}_{q}}\}+\E\{\chi_{{\cal C}_{q}}\})
\frac{\pi
                \alpha_{q} P N_{t} }{2 \sqrt{\alpha_{q}PN_{t}+1}}[{\cal U}(2^{B/Q},a/2)+{\cal V}_{M}(2^{B/Q},a/2) ].
$
 \space It, therefore, follows that 
\begin{IEEEeqnarray}{rCl}\label{eq_A1_minus_A2_l1}
        A_{1}-A_{2}&\leq&  \E\big\{(1-\chi_{{\cal C}_{q}})\}\E\{\log\big(
        1+P\vert\hat\hvec^{\dagger}_q\pvec_q
        \vert^2\big)\big\} 
        \\&& +
\frac{\pi
                \alpha_{q} P N_{t} }{ \sqrt{\alpha_{q}PN_{t}+1}}\big[{\cal U}(2^{B/Q},a/2)+
        {\cal V}_{M}(2^{B/Q},a/2)\big]
        \label{eq_A1_minus_A2_l2}
\end{IEEEeqnarray}
wherein \eqref{eq_A1_minus_A2_l1} we used the fact that $\vert\hat\hvec^{\dagger}_q\pvec_q
\vert^2$ and $\vert\hvec^{\dagger}_q\pvec^{\star}_q
\vert^2 $ are identically distributed and the Cauchy-Schwarz inequality and in \eqref{eq_A1_minus_A2_l2} we used $0\leq\E\{\chi_{{\cal A}_{q}}\},\E\{\chi_{{\cal C}_{q}}\}\leq 1$. 

It remains to bound $\E\{ 1-\chi_{ {\mathcal{C}}_q}\}$; i.e., ${\mathbb P}({\cal C}_{q}^{\rm c})$ (cf. \eqref{ep_bound_on_E_2}), where ${\mathbb P}(\cdot)$ is the probability
measure. 
Beginning with ${\mathbb P}({\cal A}_{q}^{\rm c})$, we derive a bound on  $\cos^{2}(\xi _q)$. Consider, 
$\hat{{\pvec}}_q=\cos(\phi_{q}){{\pvec}}_q +
\sin(\phi_{q})\gvec_q,$
and note that  $\pvec_q$ (cf. \eqref{eq_ZFP}) can be seen as the projection  of  ${\hat{\bar{\hvec}}}_q$  into the orthogonal complement of  $\mathcal{H}_{\mbox{\tiny-}q}= \rm{span} ({\hat{\bar{\hvec}}}_1,\ldots,$ ${\hat{\bar{\hvec}}}_{q-1},$${\hat{\bar{\hvec}}}_{q+1},\ldots,$${\hat{\bar{\hvec}}}_Q)$; after being normalized; that is,  
${{\pvec}}_q=\frac{{\Pmat}_{\mathcal{H}_{\mbox{\tiny-}q}}^\bot{\hat{\bar{\hvec}}}_q }{\| {\Pmat}_{\mathcal{H}_{\mbox{\tiny-}q}}^{\bot}{\hat{\bar{\hvec}}}_q \| }$. We further  express $\hat{\bar\hvec}_{q}$ as
$ \hat{\bar\hvec}_q=V_{1}  \pvec _{q}+V_{2} \frac{\Pmat_{\mathcal{H}_{\mbox{\tiny-}q}}{\hat{\bar{\hvec}}}_q }{\|{\Pmat}_{\mathcal{H}_{\mbox{\tiny-}q}
                {\hat{\bar{\hvec}}}_q }\|}, $
 where   $V_{1}\sim\ \Gamma (N_{\rm t}-Q+1,1)$,  $V_{2}\sim\Gamma (Q-1,1)$ and denote $W_q=\cos^{2}(\angle\langle
\hat{\bar\hvec}_q,\pvec_q\rangle)$. It, therefore, follows that   $\sqrt{W_{q}}=\frac{V_{1}^2}{V_{1}^2+V_{2}^2}\sim{\beta}(M N_{\rm t}-Q+1,Q-1)$ \cite{Roh2006}. Combining the latter representations of $\hat\pvec_{q}$ and $\hat{\bar\hvec}_{q}$, 
one obtains  
$\cos^{2} \big(\xi _q\big)=
| \hat{\bar\hvec}_q^{\dagger }\hat{\pvec}_q\big|^{2} =\big| \sin(\phi_{q})
\hat{\bar\hvec}_q^{\dagger }\gvec_q +\cos(\phi_{q})\hat{\bar\hvec}_q^{\dagger
}{{\pvec}}_q \big\vert^{2}
\geq
-\big|{\sin(\phi_{q}) \hat{\bar{\hvec}}}_q{}^{\dagger
}\gvec_q \big| ^{2}+\big| \cos(\phi_{q}) \sqrt{W_{q}}\big|^{2}
\geq\cos^{2}(\phi_{q}){W_{q}}
-\sin^{2}(\phi_{q}) 
$

Now to ${\mathbb P}\big(\mathcal{A}_q\big)
$. Consider $ {\mathbb P}\big({\mathcal{A}}_q^{\rm
        c}\big)={\mathbb P}\big(\xi _q> \frac{\pi }{2}-\theta _q\big)={\mathbb P}\big(\cos
^2\big(\xi _q\big)\leq \sin ^2\big(\theta _q\big)\big)$, then
$
        {\mathbb P}\big({\mathcal{A}}_q^{\rm c}\big)
        \leq  {\mathbb P}\big( \cos^{2}(\phi_{q})W_{q}-\sin^{2}{(\phi_{q})} \leq \sin ^2(\theta _q)\big)
        \leq 
        {\mathbb P}\big(W_q\leq \sin ^2(\theta _q)+2 \sin ^2(\phi
        _q)\big)
,$
 wherein the latter we used $\cos^{2}(\phi_{q})=1-\sin^{2}(\phi_{q})$ and $0\leq W_{q}\leq 1$. 
Let  $Y_{q}= \sin^2(\theta _q)+2 \sin ^2(\phi
_q)$ and let $F_{Y_{q}[y]}={\mathbb P}(Y_{q}\leq y)$, then,
${\mathbb P}\big({\mathcal{A}}_q^{\rm c}\big)
\leq {\mathbb P}\big(W_{q}\leq
Y_{q}\big)=\int_{} {\mathbb P}(W_{q}\leq y|Y_{q}=y)dF_{Y_{q}}(y)=\E\{{\mathbb P}\big(W_{q}\leq y|Y_{q}=y\big)\}
$. Since $W_{q}$ is independent of $\theta_{q}$ and $\phi_{q}$, ${\mathbb P}\big(W_{q}\leq y|Y_{q}=y\big)=I_y(M {N_t}-Q+1,Q-1)$
where $I_{y}$ is the regularized beta function. Using $I_y(M {N_t}-Q+1,Q-1)\leq y$ for $0\leq y\leq 1$, one obtains 
$
{\mathbb P}\big({\mathcal{A}}_q^{\rm c}\big)\leq \E\{Y_{q}\}=\E\{\sin ^2\big(\theta _q\big)+2 \sin ^2\big(\phi
_q\big)\}$.  Next, similar to \cref{Propbound on theta_q}, it can be shown that ${\mathbb P}({\mathcal{A}}_q^{\rm c})\leq 3 \big({\cal U}(2^{B/Q},a)+{\cal
V}_{M}(2^{B/Q},a)\big)$, 
where  $a, {\cal U}$ and ${\cal V}_{M}$ are defined in \cref{Theorem:Theorem1}.  To complete the proof,
we bound  
$ \mathbb{P}\big({\cal
        C}_q^{\rm c}\big)$ as follows  $ \mathbb{P}\big({\cal C}_q^{\rm c}\big)=\mathbb{P}\big((\mathcal{A}_q\bigcup\mathcal{B}_q)^c\big)\leq
\mathbb{P}(\mathcal{A}_q^c\bigcup\mathcal{B}_q^c)\leq
\mathbb{P}\big(\mathcal{A}_q^c\big)+\mathbb{P}\big(\mathcal{B}_q^c\big).$  Finally, similar to ${\mathbb
        P}({\cal A}_{q}^{\rm c})$, it can be shown that ${\mathbb P}\big({\mathcal{B}}_q^{\rm c}\big)\leq 3 (({\cal U}(2^{B/Q},a)+{\cal V}_{M}(2^{B/Q},a)\big)$, which  establishes the desired result.

\section{}
\label{App_C_ProofOLemma_7_Bound_A3_P_Q}
The proof of Lemma \ref{Lemma_Bound_on_A3_P_a_Q}  is similar to the proof of Lemma  \ref{Lemma_Lemma2_BoundonC} (cf.
Appendix \ref{App:proof
        of Lemma2:Bound
        on C}). It is obtained by substituting   $\tilde\hvec_q$,
$\hat{\tilde\hvec}_q$, $\tilde\pvec_j$, $\hat{\tilde\pvec}_j$, $\tilde\svec_{q,m}$,
$\tilde\gvec_{j,m}$, $\tilde P$, $B/(Q - \bar
Q)$ for $\hvec_q$, $\hat{\hvec}_q$, $\pvec_j$, $\hat{\pvec}_j$,
$\svec_{q,m}$, $\gvec_{j,m}$, $P$, $B/Q,$ respectively. Moreover, because \(\E\{ \|\hat{ \tilde{\hvec}}_{q,m}\|{}^2\}/ \alpha
_{q,m}
= {\tilde N}_{t}\) \cite{jindal2006mimo},  $N_t$, $a$ are replaced by $\tilde N_{t}, \tilde a$, respectively, and $\alpha_{q,m}$ by $1/M$,
due to    \cref{ass_equal_alphas}. The desired result then follows
similarly with few caveats as follows. 

The sums in \eqref{sum(|hat h p...|)}, \eqref{eq_ Exp(cdot)=D+E+F+G+H}, \eqref{eq_DoubleSum D}-\eqref{eq_D_equal_I+J}, \eqref{eq_E_equals_standard}, \eqref{AppendixA_eq16ThirdLilne}, \eqref{appendixA_eq22}
now run over ${\cal M}_{q,j}$, (cf. \cref{definitions_of_m_q_and_m_q_j}), rather than ${\cal M}$; that is, $\sum_{m=1}^{M}(\cdot)$
is replaced with $\sum_{m \in{\cal M}_{q,j}}(\cdot)$. Similarly, the double sums  in 
\eqref{eq_Xi=E(sum_sum))}, \eqref{eq_DoubleSum D}, \eqref{eq_D_equal_I+JLine1}, \eqref{appendixA_eq16SecondLine}, \eqref{double_sum_with_total_expectation} now run over  $m,n\in{\cal M}_{q,j}$.  
Furthermore,  because \(\E \left\{\|\tilde
\pvec_{j,m}\|^2
\right\} = \frac{1}{M_{j}}\), the term  $M$ in 
\eqref{eq_bound_I_CS}, \eqref{eq_E_inequality_second_standard}  is now changed to $M_{j}$.  These modifications, and \cref{ass_equal_alphas} imply that   $\sum_{m\in {\cal M}_{q,j}}\alpha_{q,m}=M_{q,j}/M,$ which thus affect \cref{PropBoundOnD,Prop_Bound_on_E_F,PropBoundOnG} in which  $\alpha_{q}/M$ is replaced with $ M_{q,j}/M_{j}M$.

Now to $                       \tilde A_{1}-\tilde A_{2}
\leq  \Delta \bar{ \tilde R}_{2,q}$; the proof is similar to that of Lemma \ref{Lemma:Lemma3 Bound A-B} (cf. Appendix \ref{AppendixProofLemma3}). It is obtained by substituting $\tilde\hvec_q$, $\hat{\tilde\hvec}_q$, $\tilde\pvec_j$, $\hat{\tilde\pvec}_j$, $M_{q}$, $\tilde P$, ${\tilde N}_{t}$, $\tilde a$, $ B/(Q - \bar Q)$ for $\hvec_q$, $\hat{\hvec}_q$, $\pvec_j$, $\hat{\pvec}_j$, $M$, $P$,  $N_t$, $a$, $B/Q,$ respectively,  in  Appendix \ref{AppendixProofLemma3}, and following similar steps.
 \end{appendices}
\begin{small}
\bibliographystyle{ieeetr}

\end{small}


\newpage
\setcounter{page}{1}
\setcounter{equation}{0}
\renewcommand{\theequation}{S.\arabic{equation}}
\setcounter{section}{0}
\setcounter{figure}{0}
\renewcommand{\thefigure}{L. \arabic{figure}}
\renewcommand{\thesection}{Supplemantary  \arabic{section}}
\renewcommand{\thesection}{\Roman{section}}
\section*{\Large{Supplementary Material}}
 \section{Proof of \cref{eq_define_DeltaR_2PandQ_simple_case}}
\label{sup_proof_cor_P_Q_symmetric}
We first show with showing that
\begin{IEEEeqnarray}{rCl}
	\label{eq_M_q}
	M_{q}=(1-\bar Q/Q)M,
	\forall q\in{\cal Q}
\end{IEEEeqnarray} and
\begin{IEEEeqnarray}{rCl}
	\label{eq_sum_over_M_q_j}
	\sum_{j\in{\cal Q}_{-q}}\frac{{M_{q,j}}}{M_j}= {(Q -\bar Q-1)}.
\end{IEEEeqnarray}  
Define 
$L\define Q/M\in \nat$, $r\define \bar
Q/Q$ and note that $r M\in \nat$.  Without loss of generality, consider  $q=1$ and assume that ${\cal Q}_{m}=\{(m-1)L+1,\ldots,m L\}$ for $m\in{\cal M}$.  
From   \cref{def_symm_policy},   \gls{ms}-1 is not served by \gls{srrh}-1, $\ldots$ \gls{srrh}-$M r$, therefore     $ M_{1}
=M(1-r)$ which establishes \eqref{eq_M_q}. To show \eqref{eq_sum_over_M_q_j}, we note that
$\forall j\in{\cal Q}_{1}$,   
\gls{ms}-{$j$}  is served by the same set of \glspl{srrh} as
\gls{ms}-1; thus, $M_{1,j}=M_{1},\; \forall j\in {\cal Q}_{1}$.   Assume, for now, that  $0\leq r\leq \frac{1}{2}$;
then by the symmetric selection policy
$M_{1,j}=M_{1}-k  $ $\forall j\in\{ {\cal
	Q}_{(1+k)\bmod M}\cup{\cal Q}_{(1-k)\bmod M}: k=1,\ldots,Mr\}$ and $M_{1,j}=M_{1}-Mr$ $\forall j\in\{{\cal Q}_{(k\bmod  M)+1}\cup{\cal Q}_{(-k\bmod M)+1}: k=Mr+1,\ldots,\lfloor M/2\rfloor
\}$. 
Thus
\begin{IEEEeqnarray}{rCl}\label{eq_sum_r_leq_half}
	\sum _{i=2}^{Q} M_{1,i}&=&(L-1) M_1+L[( M - 2 M r-1)(M_{1}-Mr)+ M r (-M r+2
	M_1-1)].
\end{IEEEeqnarray} Now to the case where   $r>\frac{1}{2}$. Here
$M_{1,j}=M_{1},\; \forall j\in {\cal Q}_{1}$, \begin{IEEEeqnarray}{rCl}
	M_{1,j}=M_{1}-k \; \forall j\in\{ {\cal
		Q}_{(k\bmod M)+1}\cup{\cal Q}_{(-k\bmod M)+1}: k=1,\ldots,M_{1}-1\} 
\end{IEEEeqnarray}
and $M_{1,j}=0$ otherwise. Thus, 
\begin{IEEEeqnarray}{rCl}\label{eq_sum_r_geq_half}
	\sum _{i=1}^{Q} M_{1,i}=(L-1) M_1+L\left(M_1-1\right) M_1.
\end{IEEEeqnarray} Substituting  $L= {Q}/{M},$ $M_{1}=M(1-r)$ and $r=\bar Q/Q$ in both \eqref{eq_sum_r_leq_half}, \eqref{eq_sum_r_geq_half}  it follows that
the two expressions are identical and are given by 
\begin{IEEEeqnarray}{rCl}
	\sum _{i=2}^{Q} M_{1,i}=\frac{M \big(Q-\bar{Q}-1\big) \big(Q-\bar{Q}\big)}{Q}
\end{IEEEeqnarray}  
which establishes \eqref{eq_M_q} and \eqref{eq_sum_over_M_q_j}.
It remains to calculate  $\tilde{T}_q=M_q \tilde{N}_t-\tilde{Q}_q+1$. Recalling that  $M_q=M (1-r)$, it is sufficient  to calculate  $\tilde{\mathcal{Q}}_q$, which is given by 
\begin{IEEEeqnarray}{rCl}
	\tilde{Q}_q=Q-\sum_{j\in\mathcal{Q}_{\mbox{\tiny-}q}} \chi _{\{0\}}\left(M_{q,j}\right)
\end{IEEEeqnarray} Note that  $\sum_{j\in\mathcal{Q}_{\mbox{\tiny-}q}} \chi _{\{0\}}\left(M_{q,j}\right)$  can be written as  $\left| \left\{j,\left| \mathcal{M}_{q,j}\right| =0,j\neq q\right\}\right|$; i.e, the number of \glspl{ms}  served by at least one \gls{srrh} that serves \gls{ms}-$q$. For $r\geq \frac{1}{2}$,   the latter sum is equal to $2  Q(1-r)-L-1$,whereas for $r<1/2$, $\tilde Q_{q}=Q-1$.  
This
can be written as, $\tilde{Q}_q=Q-\min \big(Q,-2 \bar{Q}+\big(2-{1}/{M}\big) Q-1\big)$, and therefore 
\begin{IEEEeqnarray}{rCl}\tilde{T}_q=M (1-r) \tilde{N}_t+1-\min \left[Q,\left(2-\frac{1}{M}\right) Q-2 \bar{Q}\right]
\end{IEEEeqnarray}
which establishes the desired result.
\section{A Detailed Description of the Simulation Setup}
\label{Sec:DetaliedSimulationSetup} 
We now provide a detailed description of the two simulation setups considered in \cref{FigSystemSimulation}
\label{SubSecRandomlyDistributedMS}
Here we consider a practically oriented setup with randomly dispersed \glspl{ms} while considering propagation loss and shadowing. 
The setup includes a cluster of \(M = 4\) \glspl{srrh}, each placed at the center of one of four adjacent hexagons, creating a hexagonal grid with an edge-length of 100 m; i.e., we cut four hexagons and placed an \gls{srrh} at the center of each
hexagon. Each \gls{srrh}  is equipped with \(N_t = 8\)
isotropic transmit antennas. Eight  single-antenna \glspl{ms}  \((Q = 8)\)  were  placed uniformly at random
in the area spanned by the hexagons, with a minimum distance of 10 m between each \gls{ms} and \gls{srrh}. The results were averaged over 20 realizations of \gls{ms}-placements, where each realization determined a set of attenuation factors \(\alphavec=\{\alpha_{q,m}:q=1\cdots8, m=1\cdots 4\}\) according     
\(\alpha_{q,m}=-128-37.6\log_{10}(r_{q,m})\) (in dB),\footnote{This model was used for urban-area non-line-of-site links  by the 3GPP; cf. page 61 3GPP
	Technical Report 36.814 \cite{attunuation2010}.} where  \(r_{q,m}\) is the distance
from S-RRH-\(m\) to MS-\(q\) in Km. The noise level at the receivers was  \(-121\) dBm.
For  each realization $\alphavec$, we calculated the rates $\hat R_{q}(\alphavec)$, $\hat{\tilde R}_{q}(\alphavec)$ using \eqref{eq_TrueRate} and \cref{Def_P_and_Q_Rate}, for the corresponding scheme,\footnote{$\hat R_{q}(\alpha)$
	is a function of $\alphavec$ by \eqref{eq_SignalModelq}
	and by \cref{ass_channelModel}.} 
by averaging  over   40  realizations of $\{ \hvec_{q,m}\}_{q\in {\cal Q},m\in{\cal M}}$, generated according to \cref{ass_channelModel}. The network throughput for a given $\alphavec$ 
was calculated as $\bar R(\alphavec)=1/Q \sum_{q=1}^{Q}\hat R_{q}(\alphavec)$, and the  throughput $\bar R$ was calculated by  
averaging  \(\bar R(\alphavec)\) over  20 realizations of \(\alphavec\); i.e., $\bar R =1/20\sum_{i=1}^{20} \bar R(\alphavec_{i}),$ where  each \(\alphavec_{i}\) corresponded to a different  \gls{ms} placement. \(\bar{\tilde R}(\alphavec)\) and \(\bar{\tilde R}\) are defined similarly.

To ensure a fair comparison, in all figures (\cref{FigSystemSimulationA,FigSystemSimulationB}) we considered that the transmitters could always turn off some of their antennas to reduce the effective \gls{miso} channel dimensions.
Therefore, in the standard scheme, for
each $\alphavec$, we evaluated $\bar R(\alphavec)$
for \({N}_{t}= \{2,\ldots,8\}\) and picked the maximum. We applied a similar procedure for the \gls{paq} scheme, where we maximized the rate over \(\bar Q\), with $  N_{t}= 8$; i.e.,  by evaluating $\bar {\tilde R}(\alphavec)$ for $\bar Q = \{1,\ldots,6\}$ and  taking the maximum.

Finally, we used the following power allocation strategy.
\begin{definition}[The equal power-backoff  strategy]
	\label{Def_power_standard_shceme}
	Considering the standard feedback scheme,  each  \gls{srrh} transmits with equal power to each \gls{ms}; i.e., $P_{q,m}=P,
	\forall q\in \cal{Q}\), \(
	m\in \cal{M}$. To avoid violating its individual power constraint \(P_{\rm max}\),  each  \gls{srrh}  sets   \(P=\frac{\gamma P_{\rm max}}{Q}\), where  \(\gamma\geq 1\) is
	the backoff factor, given by
	\begin{IEEEeqnarray}{rcl}
		\label{eq_define_gamma}
		\gamma=\min\Big\{\gamma_{m}:\sum_{q\in{\cal Q}}\gamma_m\Vert
		\pvec_{q,m}\Vert^2=Q,\;m\in{\cal M\Big\}}
	\end{IEEEeqnarray}
	To explain this strategy, we note that  \(\E\{\|\xvec_m \|^2\vert U\} = \sum _{q\in {\cal Q}}\frac{P_{\rm max}}{Q}\Vert \pvec_{q,m} \Vert^2 \leq P_{\rm max}\), and therefore
	\(\sum _{q\in {\cal Q}}\Vert \pvec_{q,m} \Vert^2 \leq Q , \forall m\in\cal M\). Because  \(\Vert \pvec_{q} \Vert = 1\), the latter constraint is always satisfied with strong inequality, implying that it is possible to increase the
	power.  The objective of  \eqref{eq_define_gamma} is to guarantee that at least one \gls{srrh} transmits with maximal power.\footnote{Note that while this power allocation strategy is not optimal, it yields good performance in high \glspl{snr}.} 
	Considering the \gls{paq} scheme, the strategy in \cref{Def_power_standard_shceme} applies with a minor modification. Each \gls{srrh} serves exactly \(Q -\bar
	Q\) \glspl{ms}, with power \(\tilde P_{q,m}=\tilde P=\frac{\tilde\gamma P_{\rm max}}{Q -\bar
		Q }, \forall q\in \cal{Q}\), \( m\in \cal{M}$, where \(\tilde\gamma\)
	is defined similarly to \eqref{eq_define_gamma} while substituting
	$\tilde \pvec_{q,m}$ for $\pvec_{q,m}$ and $Q-\bar Q$ for $Q$.
\end{definition}

Fig. \ref{FigSystemSimulationA} presents
the throughput as a function of each \gls{srrh} transmit power, \(P_{\rm max}\) (cf. \cref{ass_power_constraint}).
\cref{FigSystemSimulationB} presents the average throughput as a function of $B$,
under a per-\gls{srrh} power constraint
of   \(P_{\rm max}=\) \(45\) dBm. 


\end{document}